\documentclass{emulateapj}
\usepackage{natbib}
\usepackage{apjfonts}
\usepackage{epsfig}

\newcommand{\etal}{et al.}

\def\yr{{\rm\thinspace yr}}
\def\Msun{\hbox{$\rm\thinspace M_{\odot}$}}
\def\Msunpyr{\hbox{$\Msun\yr^{-1}\,$}}

\def\ltsima{$\; \buildrel < \over \sim \;$}
\def\simlt{\lower.5ex\hbox{\ltsima}}
\def\gtsima{$\; \buildrel > \over \sim \;$}
\def\simgt{\lower.5ex\hbox{\gtsima}}

\def\<{\thinspace}

\def\msun{{\rm M_{\odot}}}

\shorttitle{Early growth of massive black holes}
\shortauthors{Pelupessy \etal}
\begin{document}

\title{How rapidly do supermassive black hole ``seeds'' grow at early times?}
\author{Federico I. Pelupessy\altaffilmark{1}, 
Tiziana Di Matteo\altaffilmark{1}, Benedetta Ciardi\altaffilmark{2}}
\altaffiltext{1} {Physics Department, Carnegie Mellon University, 5000 Forbes
  Avenue, Pittsburgh, PA 15213}
\altaffiltext{2}{Max-Planck-Institut f\"{u}r Astrophysik, 
Karl-Schwarzchild-Strasse 1, 85740 Garching bei M\"{u}nchen, Germany}

\begin{abstract}

We investigate the physical conditions for the growth of intermediate mass
seed black holes assumed to have formed from remnants of the first
generation of massive stars. We follow the collapse of high-$\sigma$ halos
with $T_{vir} > 10^4$ K using cosmological, smooth-particle hydrodynamic
(SPH) simulations in the standard $\Lambda$CDM model. During collapse of the
parent halo the seed holes are incorporated through mergers into larger
systems and accrete mass from the surrounding gas. We include a
self-consistent treatment of star formation, black hole accretion and
associated feedback processes.  Even under optimistic assumptions for the
seed black hole mass and for efficient merger rates, we find that seed holes
in halos $M \le 10^{10} \Msun$ never reach the conditions for critical
Eddington growth.  Most of the black hole growth in this regime is
determined by the initial mass and the merger rates. Critical accretion
rates are reached, albeit only after a significant delay, at the time of
collapse ($z\sim 7$) for $3$-$4\sigma$ halos of $M\sim 10^{11}\Msun$.  Our
results imply $M_{BH} = 5\times 10^6 \Msun (M_{halo}/10^{11}\Msun)^{0.78}$
at the time of collapse. The required conditions of  Eddington growth to
explain the build-up of supermassive black holes ($\sim 10^9 \Msun$), as
implied by Sloan quasars at $z >6$, are therefore hard to meet in such a
scenario. Without a 'jump-start' these conditions may be only achieved in
extremely rare halos with $M_{halo} > 10^{13}$ that collapsed before $z \sim
6$. The sub-Eddington regime in which black holes holes accrete at early
time implies a small contribution to the reionization by miniquasar but
still sufficient to cause appreciable heating of the IGM at $z \lesssim
15-18$.

\end{abstract}

\keywords{quasars: general --- galaxies: formation --- galaxies: active --- 
galaxies: evolution --- cosmology: theory --- hydrodynamics}

\section{Introduction}
\label{sec:intro}

Following the discovery of quasars \citep{Schmidt1963, Greenstein1963} it was
suggested that supermassive black holes ($10^6-10^9$~\Msun) lie at the centers
of galaxies, and that the quasar activity is fueled by the release of
gravitational energy from their accreted matter. The remnants of quasar phases
at early times are probably the supermassive black holes (SMBHs) found at the
centers of galaxies in our local Universe. Indeed, the properties of SMBHs
found at the centers of galaxies today are tightly coupled to those of their
host galaxies thus providing strong observational evidence for a close
connection between the formation and evolution of galaxies and of their
central black holes \citep{Magorrian1998, Ferrarese2000,
Gebhardt2000}. Remarkably, quasars with inferred black hole masses in excess
of $10^9$~\Msun\ have now been discovered out to $z\sim 6$ (Fan et al. 2003),
posing a significant challenge for theoretical models of high-redshift quasar
and galaxy formation. The origin, seed mass and the physical conditions
for their subsequent growth to supermassive black holes remain uncertain.

The various scenarios that have been proposed for the black hole `seed'
progenitors can be broadly grouped into two main ideas.  One possible route
traces the black hole 'seeds' to the remnants of the first generation of
PopIII stars.  Numerical simulations of the fragmentation of primordial clouds
in the standard cold dark matter model suggest that these PopIII, formed of
metal-free gas at $z\sim 20-30$, are indeed very massive~\citep{Bromm1999,
Abel2000, Nakamura2001, Yoshida2003, Gao2006} 
and are expected to leave behind a
remnant black hole seed in the range of $10-10^4 \Msun$.  Alternatively, a
different mass scale of $10^5 - 10^6 \Msun$ is predicted from models in which
seed black holes form directly from the collapse of low angular momentum,
dense gas in the center of massive halos with virial temperatures above $10^4
$~K~\citep{HaehneltRees1993, Umemura1993, LoebRasio1994, Eisenstein1995,
Bromm2003, Koushiappas2004, Begelman2006}. Even though the disposal of angular
momentum from the gas is a strong constraint in these models, they offer a
promising scenario: seed black holes formed in this way have a `jump-start'
for their growth into the Sloan quasars.

In the most commonly adopted scenario, in which the first black holes are
traced to the first generation of stars, growing the seeds up to $~10^9
\Msun$, (as implied by the Sloan quasars at $z\sim 6$) requires an almost
continuous accretion of gas at the critical, Eddington rate. If we assume that
early black holes grow by accretion at some fraction $f$ of the Eddington rate,
then the black hole mass $M_{\rm BH}(z) = M_0 \exp(f \Delta t(z,z_0)/t_{s})$,
with $t_{s}$ the Salpeter time and $M_0$ and $z_0$ are the formation redshift
and mass, while $\Delta t(z,z_0)$ indicates the time between $z_0$ and
redshift $z$. From this follows that with less than about $\sim 800$ Myr to
$z=6$ and a Salpeter time of $t_s=45$ Myr (assuming a radiative efficiency of
10\%), the accretion rate should be $f > t_s/{\Delta t} \;\log(M_{(z=6)}/M_0)
\sim 0.6 -0.9$ of Eddington, or equivalently an Eddington accretion rate
should be sustained at least for $f\sim 60 - 90 \%$ of the time to $z =6$. It
is an open question however whether such vigorous Eddington growth can be
reached soon enough and sustained in the shallow potential at early times.
The question of how much mass is accreted and therefore how much energy is
liberated by the first miniquasars has also a significant impact for the
proposed scenarios of reionization and X-ray heating of the intergalactic
medium (IGM)\citep[e.g.][]{Madau1999, Valageas1999, Miralda2000, Oh2001,
Venkatesan2001, Wyithe2003c, Madau2004, Ricotti2004, Zaroubi2006}. 
The first black holes can
potentially produce very intense ionizing radiation and due to their X-ray
radiation are efficient at heating up the surrounding IGM.

In this paper, we use high resolution SPH simulations of massive, isolated
(3-4 $\sigma$) halos to study the early growth and evolution of the first
seed black holes and explore the impact of their evolution to reionization
and X-ray heating. Semi-analytic models that consider plausible scenarios
for the hierarchical assembly and growth of massive black holes from such
seeds \citep{Volonteri2003, Volonteri2005a, Madau2004} have either assumed
Eddington accretion or have argued that quasi-spherical, 'Bondi' accretion
in halos with $T_{vir} > 10^4$ K, where efficient cooling via hydrogen
atomic lines could occur, can sustain critical and short phases of 
supercritical accretion \citep{Volonteri2005b}.  Multi-scale simulations
that follow the hierarchical assembly history in $\Lambda$CDM of the most
massive halos forming in a $\sim 3 $ Gpc$^3$ volume at $z\sim 6.5$ as well
as the associated black hole growth have been able to show such rare massive
halos are good candidates for the hosts of the first quasars \citep{Li2006}.
Here, we follow these earlier papers in considering the hierarchical
assembly of SMBHs and use the approach of direct SPH cosmological
simulations to consider the physical conditions of the gas inflows that feed
the central black hole in its early growth. Thanks to the faithful tracking
of the gas dynamics these provide a self-consistent treatment of the
coupling of large-scale gas inflow, black hole growth and associated
feedback processes. In particular, we will base our treatment of accretion
onto massive black hole and associated energy feedback on the method
recently developed by \citet{DiMatteo2005, Springel2005a} in fully
three-dimensional hydrodynamic simulations, where, for example, we were able
to explicitly confirm that self-regulated quasar growth can reproduce the
observed $M_{\rm BH}$-$\sigma$ relation. We focus on understanding the 
black hole growth history at early time and as a function of halo mass.  We
do not impose a critical Eddington growth phase, but instead attempt to
assess the detailed black hole accretion history (our work can thus be seen
as complementary to \citet{Li2006}). Our aim is to test how reliable the
assumption of Eddington growth is likely to be, in particular close to the
time of reionization.

In \S \ref{sec:sim} we review our simulation method and in particular the implementation
of the black hole accretion and feedback model, as well as our recipe for
seeding primordial minihalos with black holes. We then discuss the general
properties and evolution of the black holes in simulations of halos of various
masses (\S \ref{sec:results}). We explore the implications of our results for BH growth
scenarios as well as for the IGM reionization and heating in \S \ref{sec:disc}. 
In \S \ref{sec:concl} we briefly summarize our main results. 
Throughout, a standard $\Lambda$CDM cosmological model is assumed,
with $\Omega_{\Lambda}=0.7$, $\Omega_m=0.3$, baryon density $\Omega_b=0.04$
and Hubble constant $H_0=100 h$ km/s/Mpc ($h=0.7$).

\section{The Simulation Method}
\label{sec:sim}

\subsection{SPH cosmological simulations}
The simulations are performed with the parallel cosmological TreePM-Smooth
Particle Hydrodynamics (SPH) code {\small GADGET2}~\citep{Springel2005d}.
This code has been extensively tested in a wide range of applications from
large scale structure formation to star formation. For the computation of the
gravitational force, the code uses the Tree-PM method that combines a 'tree'
algorithm for short range forces and a Fourier transform particle-mesh method
for long range forces. {\small GADGET2}'s implementation of SPH~\citep[e.g]
[]{Monaghan1992} uses a formulation which manifestly conserves energy and
entropy while using fully adaptive SPH smoothing lengths~\citep{Springel2002}.
Both the force computation and the time stepping of the code are fully
adaptive. Radiative cooling and heating process are included \citep[as
in][]{Katz1996}, as is photoheating due to an imposed ionizing UV background.
Detailed tests of the basic performance and accuracy of {\small GADGET2} for
gravitational and hydrodynamic problems are presented in~\citet{Springel2005b}.

The interstellar medium (ISM), star formation and supernovae feedback as well
as black hole accretion and associated feedback are treated by means of
sub-resolutions models. In particular, the multiphase model for star forming
gas developed by \cite{Springel2003a} has two principal ingredients: (1) a
star formation prescription and (2) an effective equation of state (EOS). For
the former we adopt a rate motivated by observations and given by the
Schmidt-Kennicutt Law \citep{Kennicutt1989}, where the star formation rate is
proportional to the density of cold clouds divided by the local dynamical time
and normalized to reproduce the star formation rates observed in isolated
spiral galaxies~\citep{Kennicutt1989, Kennicutt1998}. The effective EOS 
encapsulates the self-regulated nature of star formation due to supernovae
feedback in a simple model for a multiphase ISM. In this model, a thermal
instability is assumed to operate above a critical density threshold
$\rho_{\rm th}$, producing a two phase medium consisting of cold clouds
embedded in a tenuous gas at pressure equilibrium. Stars form from the cold
clouds, and short-lived stars supply an energy of $10^{51}\,{\rm ergs}$ to the
surrounding gas as supernovae. This energy heats the diffuse phase of the ISM
and evaporates cold clouds, thereby establishing a self-regulation cycle for
star formation.  $\rho_{\rm th}$ is determined self-consistently in the model
by requiring that the EOS is continuous at the onset of star formation. The
cloud evaporation process and the cooling function of the gas then determine
the temperatures and the mass fractions of the two 'hot and cold' phases of
the ISM, such that the EOS of the model can be directly computed as a function
of density.  The approach we adopt here for star formation (and the parameters
we use) has already been shown to lead to a numerically converged estimate for
the cosmic star formation history of the universe that agrees well with low
redshift observations \citep{Springel2003b}.

A prescription for accretion and feedback from massive black holes is also
included~\citep{DiMatteo2005, Springel2005a}. Technically, we represent black
holes by collisionless particles that grow in mass by accreting gas from their
environments. A fraction of the radiative energy released by the accreted
material is assumed to couple thermally to nearby gas and influence its motion
and thermodynamic state. Our underlying assumption is that the large-scale 
feeding of galactic nuclei with gas (which is resolved in our simulations) is
ultimately the critical process that determines the growth of massive black
holes. Here we briefly review the main features and ingredients of this model.

\begin{table}[t]
\begin{center}
\caption{ Overview of the simulation parameters of the main runs. The runs are
labeled with a letter indicating the type of run, and a number indicating the
mass.}
\label{tab:runs}
\begin{tabular}{c c c c c c}
\hline \hline\\
label \ \ & mass ($\Msun/h$) \ \ & $z_{\rm vir}$ \ \ \ &$\lambda $ \ \ \ & $N$ (total,
gas) \ \ & $\epsilon_{\rm f} \ $ \\ 
\hline \\
A1 & $10^8$ & 16 & 0.03 & $10^6,10^5$ & 0.05 \\
A2 & $10^9$ & 12 & 0.03 & $10^6,10^5$ & 0.05 \\
A3 & $10^{10}$ & 10 & 0.03 & $10^6,10^5$ & 0.05 \\
A4 & $10^{11}$ & 7.5 & 0.03 & $5\times10^6,5\times10^5$ & 0.05 \\
B1 & $10^8$ & 16 & 0.03 & $10^6,10^5$ & 0 \\
B2 & $10^9$ & 12 & 0.03 & $10^6,10^5$ & 0 \\
B3 & $10^{10}$ & 10 & 0.03 & $10^6,10^5$ & 0 \\
B4 & $10^{11}$ & 7.5 & 0.03 & $5\times10^6,5\times10^5$ & 0 \\
C1 & $10^8$ & 30 & 0.03 & $10^6,10^5$ & 0.05 \\
C2 & $10^9$ & 20 & 0.03 & $10^6,10^5$ & 0.05 \\
C3 & $10^{10}$ & 16 & 0.03 & $10^6,10^5$ & 0.05 \\
C4 & $10^{11}$ & 9 & 0.03 & $10^6,10^5$ & 0.05 \\
\hline
\end{tabular}
\end{center}
\end{table}

\subsubsection{The seed black hole}
There is no clear consensus for the mass of the first black holes. A number of
scenarios exist which suggest that rather than starting out as stellar mass
black holes, the first black holes are formed as intermediate mass black
holes. ``Seed'' black holes can form as the end-point of the evolution of 
the first generation of stars \citep[e.g][]{Bromm1999, Abel2000,
Nakamura2001, Yoshida2003, Gao2006} and in this case a BH of mass a few
$100-10^4~\msun$ can result \citep[e.g][]{MadauRees2001, Heger2003,
Ricotti2004}. The most massive Pop III stars will form in rare high density
peaks of the primordial density field \citep{Gao2006}. This will be the
starting point for our study.

We introduce collisionless `sink' particles in the simulations to model
black holes at the centers of minihalos. In order to achieve this we keep
track of the formation of minihalos by running a friends-of-friends (FOF)
group finder. For reasons of performance, and because the identity of halos
changes only slowly, the group finder is not run every timestep but at times
equally spaced in log scalefactor, with $\Delta \log{a} = \log{1.25}$. The
group finder is run with a linking length of 0.16 and we select objects
with  a mass of $10^6 \Msun$
and place a seed BH of $M_{\rm seed} = 10^4 \Msun$ in them if
they do not already contain a BH. In practice most of the halos of this mass
and their black holes are formed between $z=15$ and $z=30$. 

 Our choice of $M_{\rm seed} = 10^4 \Msun$ is on the high side for Pop III
remnants. This mass corresponds approximately to the total available reservoir
of star forming gas in these halos as found recently by \citet{Gao2006}. In
this work, the authors could not accurately predict how much of this gas would
end up in the Pop III star and although they argued that the growth probably
would stop before the stellar object reaches $10^3 \Msun$ these results are
subject to considerable uncertainty. Hence, here we have explored the most
optimistic case (from the point of view of studying early SMBH formation from
stellar seeds) in which all of the total gas reservoir of $10^4 \Msun$ has
gone into the formation of first star. In \S \ref{sec:disc} we will however also
discuss results for a more conservative value of $M_{\rm seed}=10^3
\Msun$. Alternatively, these seeds can be viewed as the end results of direct
gas collapse scenario or stellar gravitational collapse scenario (although
these formation mechanisms are somewhat more speculative).

As we shall see the black holes do not undergo much evolution at early
times, so our results are rather insensitive to the choice of the seeding
procedure.  In particular different choices for the linking length (the
over-density) give similar results, although choosing it too low tends to
select objects that have not yet completed collapse and that are extended
and non-spherical. After the seeding of the BH by this procedure, further
growth of the black hole sink particles proceeds by gas accretion, at
a rate that depends sensitively on the local gas conditions, or by mergers with
other black hole particles.

\subsubsection{Black hole accretion \& feedback model} 
The accretion model is fully described and tested in
\cite{Springel2005a, DiMatteo2005}.
We relate the (unresolved) accretion onto the black hole to the large scale
(resolved) gas distribution using a Bondi-Hoyle-Lyttleton parameterization
\citep{Bondi1952, BondiHoyle1944, Hoyle1939}.~In this description, the
accretion rate onto the black hole is given by $\dot{M}_{\rm B} \, = \,
{{4\pi \, \, [G^2 M_{\rm BH}^2 \, \rho}] / {(c_s^2 + v_{BH}^2)^{3/2}}} \, $
where $\rho$ and $c_s$ are the density and sound speed of the ISM gas
respectively, and $v_{BH}$ is the velocity of the black hole relative to the
gas. However, because of the strong dependence on sound speed of the Bondi
accretion rate, the effective $c_s$ resulting from our multiphase ISM model
will  give a poor estimate of the accretion rate. If we make the assumption
that the time spent in the cold and hot phase of the ISM is proportional to
their respective volume filling factors $f_c$ and $f_h$, then formally
\begin{eqnarray}
\label{eq:bondi_multiphase}
\dot{M}_{\rm B} & = &
\frac{4\pi  G^2 M_{\rm BH}^2 \, f_c \rho_c}{(c_c^2 + v_{BH}^2)^{3/2}}+
\frac{4\pi  G^2 M_{\rm BH}^2 \, f_h \rho_h}{(c_h^2 + v_{BH}^2)^{3/2}}
 \nonumber \\ 
 & = &
\frac{4\pi  G^2 M_{\rm BH}^2 \, x \rho}{(c_c^2 + v_{BH}^2)^{3/2}}+
\frac{4\pi  G^2 M_{\rm BH}^2 \, (1-x) \rho}{(c_h^2 + v_{BH}^2)^{3/2}}.
\\ \nonumber
\end{eqnarray}
Here the subscripts $c$ and $h$ denote quantities in the cold and hot phase,
$x$ is the mass fraction in the cold phase, $x \equiv f_c \rho_c / \rho$.
Normally the cold phase, and hence the first term in
Equation~\ref{eq:bondi_multiphase}, dominates the accretion rate. We use this
expression to calculate the accretion rates from the local conditions at the
BH positions. We also set a limit to the accretion rate by assuming the
accretion rate cannot be higher than the Eddington rate, $\dot{M}_{\rm Edd} \,
\equiv \, ({{4\pi \, G \, M_{\rm BH} \, m_{\rm p}}) / ({\epsilon_{\rm r} \,
\sigma_{\rm T} \, c}}) \, $ where $m_{\rm p}$ is the proton mass, $\sigma_{\rm
T}$ is the Thomson cross-section, and $\epsilon_{\rm r}$ is the radiative
efficiency.  The latter relates the radiated luminosity, $L_{\rm r}$, to the
accretion rate, $\dot {M}_{\rm BH}$ as $ \epsilon_{\rm r} \, = \, {{L_{\rm
r}}/({\dot {M}_{\rm BH} \, c^2}}) \,$, which simply gives the mass to energy
conversion efficiency set by the amount of energy that can be extracted from
the innermost stable orbit of an accretion disk around a black hole.  We will
adopt a fixed value of $\epsilon_{\rm r} =0.1$, which is the mean value found
for a radiatively efficient \citep{Shakura1973} accretion disk onto a
Schwarzschild black hole.

As in \cite{Springel2005a}, finally, we assume that a fraction $\epsilon_{\rm
f}$ of the radiated luminosity $L_{\rm r}$ couples to the surrounding gas in
the form of feedback energy, viz. $ \dot{E}_{\rm feed} \, = \, \epsilon_{\rm
f} \, L_{\rm r} \,\, = \, \epsilon_{\rm f} \, \epsilon_{\rm r} \, \dot{M}_{\rm
BH} \,c^2 \, $.  For simplicity, we model this energy deposition as thermal
energy deposited isotropically in the region around the black hole.  Lack of
spatial resolution precludes us from considering mechanical modes of releasing
the energy, like jets and winds. However, it is likely that any form of black
hole feedback will lead to a fraction of the energy being thermalized
eventually, and that the final impact of the feedback depends primarily on the
total amount of energy released and less on the form it is released in.

\begin{figure*}
    \centering
    \epsscale{0.49}
    \plotone{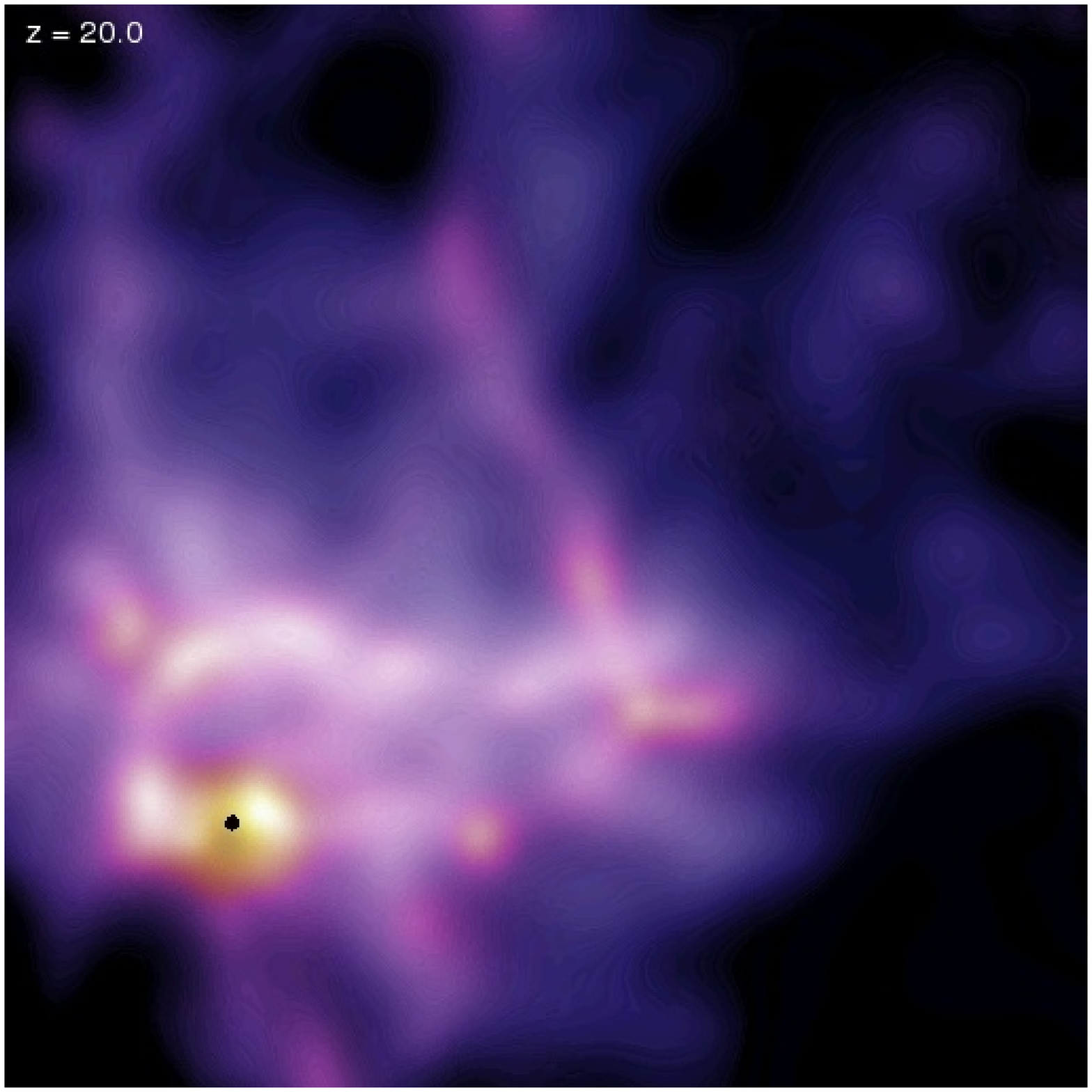}
    \plotone{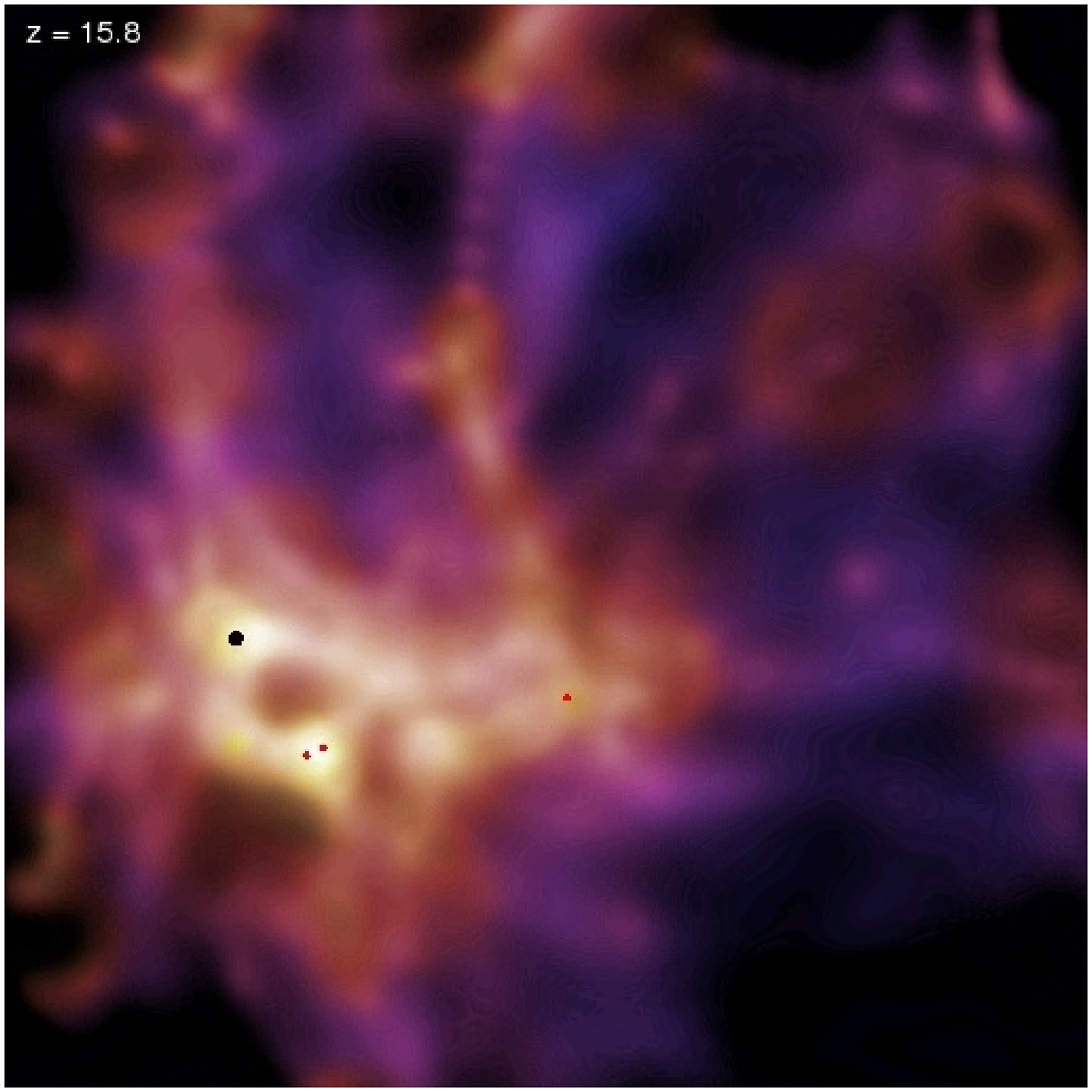}
    \plotone{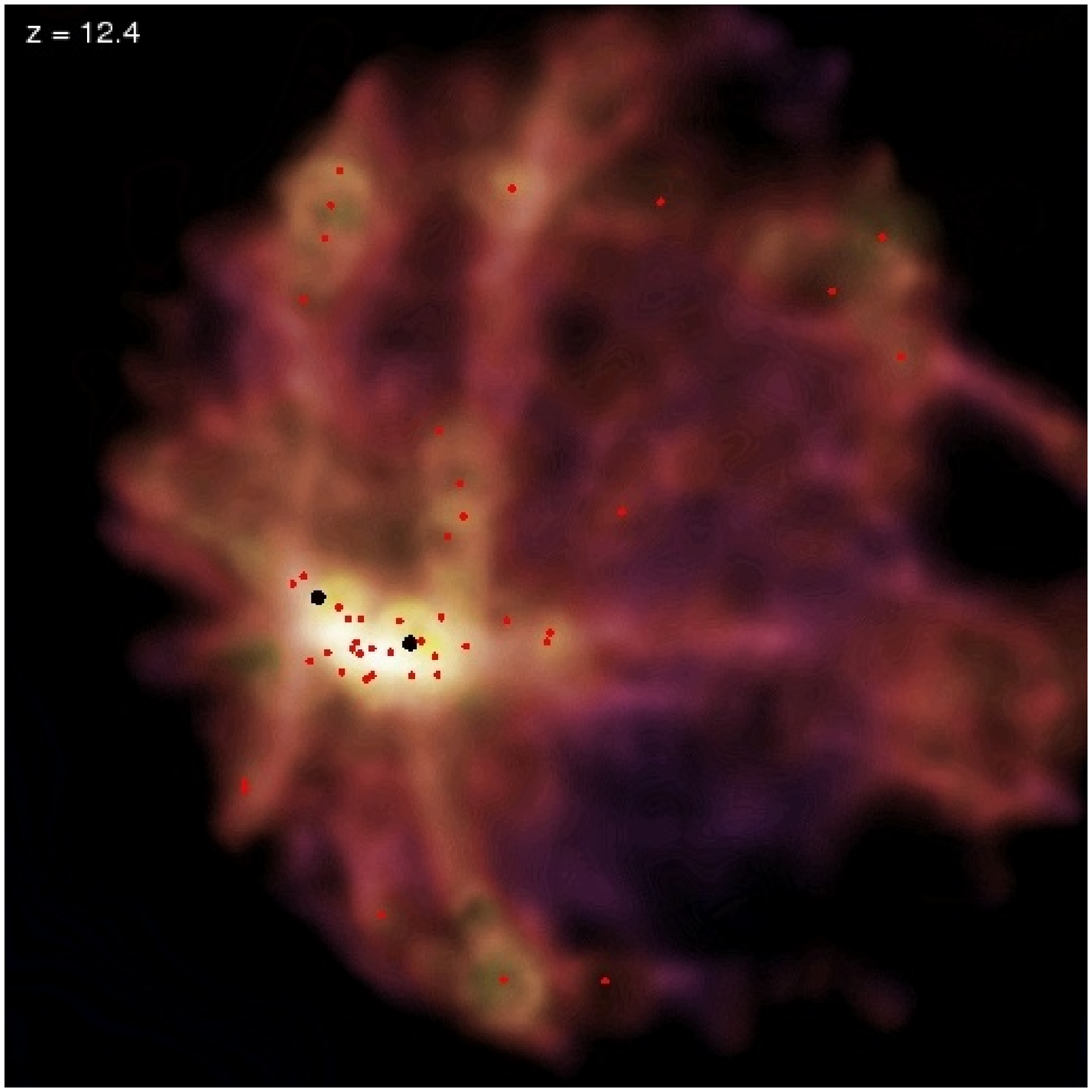}
    \plotone{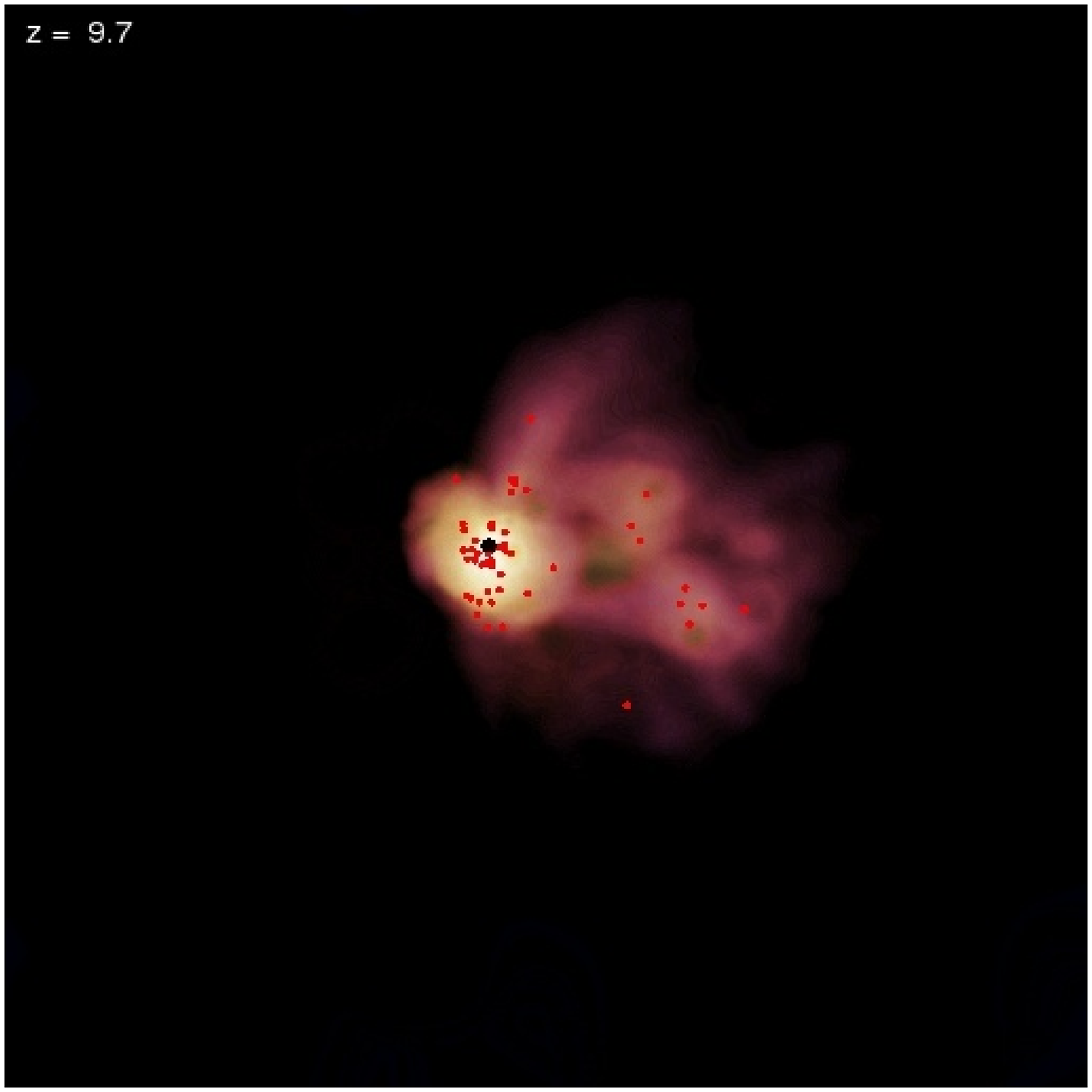}
    \caption{Evolution of gas density for a representative run. Shown is the
    gas density (indicated by brigthness) color coded by temperature for the
    A3 run at (from top left to bottom right) redshifts of 20,15.7, 12.4 and
    9.7. Width of the panels is 200 kpc (comoving).  BH twice as massive as
    the seed are marked by red dots, the largest BH in each frame is enlarged
    and shown in black (for redshift 12 the two heaviest differ less than 20\%
    in mass, and are both enlarged).}
    \label{fig:tophat}
\end{figure*}

\subsubsection{Black Hole Mergers} 
When two or more objects merge to form a single dark matter halo, dynamical
friction on their central black holes quickly brings them to the center of the
halo where they are also expected to merge eventually - hierarchical black
hole mergers contribute to the growth of the central black holes.  However,
whether black hole binaries coalesce efficiently is still a matter of
debate. In a stellar environment, it has been argued that the binary hardens
very slowly \citep{Begelman1980, Milosav2003}. In gaseous environments however
binaries do coalesce more rapidly owing to strong dynamical friction within
the gas \citep{Makino2004, Escala2004}. In our galaxy-sized simulations
\citep[e.g][] {DiMatteo2005}, and even more so in the cosmological top-hat
models, it is not possible to treat in detail the problem of binary
hardening. Because galaxies have typically large central concentration of gas
we instead assume that two black hole particles merge if they come within the
spatial resolution of the simulation (i.e. within the local SPH smoothing
length) \emph{and} their relative speed lies below the local sound
speed.  This could introduce a dependence of the derived merger rates 
on the resolution, however we will check that this is not the case
 (see \S \ref{sec:restest}).

In the final stage of the black hole mergers, the emission of gravitational
waves carries linear momentum, which can cause the black hole to recoil
\citep{Madau2004, Volonteri2006}.  Note that in our cosmological simulations
we do not have the resolution nor the relativistic physics required to
directly calculate the ejection of black holes by gravitational recoil. If the
recoil is larger than the halo escape velocity the black hole will be ejected
from its halo~\citep{Haiman2004, Yoo2004, Volonteri2006}. This effect is only
important for similar mass mergers, moreover, the proper method to calculate
the recoil velocity is still uncertain, leading to several order of magnitude
variation according to different choices of parameters. We will discuss the
possible implications of neglecting this effect in \S \ref{sec:disc}.

\subsection{Initial Conditions}
We set-up the initial conditions as isolated spherical overdensities
that correspond to high-$\sigma$ peaks in the Gaussian random field of
cosmological density fluctuations \citep{Katz1991}. In this configuration,
the `top-hat' (i.e. uniform density) cloud is given perturbations (using the
Zeldovich approximation) with a given power spectrum $P(k) \propto k^{-2.5}$
appropriate for the scales we study.  This is the same technique as that
exploited recently by \citet{Bromm2003,Bromm2004} in the treatment of
the primordial star formation problem, except that here we are interested in
larger halos ($T_{\rm vir} > 10^4$ K) which cool efficiently via atomic
hydrogen lines and for which H$_2$ cooling is not important.

Specifically, the following procedure is employed to set up the particle
positions and velocities: first, a regular grid of dark matter (DM) particles
is setup. The DM particles are assigned perturbations using the Zeldovich
approximation, which also sets the peculiar velocities. The amplitude is
normalized as in \cite{Bromm2002} by fixing the initial variance $\sigma_i$ of
the fluctuation spectrum at the starting redshift $z_i$,
\begin{equation}
\sigma^2_i = A \sum k^n,  
\end{equation}
such that
\begin{equation}
\sigma(z_{vir})=\frac{1+z_i}{1+z_{vir}} \sigma_i=1,
\end{equation}
where $z_{\rm vir}$ is the target redshift of collapse (i.e. virialization).
Particles from a spherical region of size $R=(3 M / 4 \pi \rho)^{1/3}$ are then
selected and given a Hubble expansion (modified to affect a collapse at the
target redshift). A pre-determined amount of angular momentum is imparted by
assigning rigid rotation to the particles with a spin parameter $\lambda=L
|E|^{1/2}/G M^{5/2}$,  where  $L$ is the angular momentum, $|E|$ the total
energy and $M$ the total mass. This is necessary because  we cannot follow
self-consistently the angular momentum evolution of the collapsing object.
Gas particles are setup in the same way, except that they are not subject to
the random perturbations, as on these scales pressure forces will have
erased any initial density fluctuations.

\subsection{The simulation parameters}
Using the above procedure we can investigate different collapse scenarios by
choosing the total mass $M$, the redshift of collapse $z_{\rm vir}$ and
$\lambda$. In Table~\ref{tab:runs}, we summarize the simulation parameters of
the systems we examine. They comprise halos of $M=10^{8}, 10^9, 10^{10}$ and
$10^{11} \Msun$. For $z_{\rm vir}$ we will take the collapse redshift of a
3-$\sigma$ peak (for $\sigma_8=0.9$, for a $\sigma_8=0.75$ these would be
$\approx 3.3$-$\sigma$) of the corresponding mass, i.e. $z_{\rm vir} = 16,12,
10$ and $7.5$ respectively for these halos. Note, however, that in dense
regions, likely the precursors of the halos that harbor high redshift quasars,
high density peaks are over-represented, as found e.g. in recent numerical
investigations by \citet{Gao2005} (where for example the precursor of
simulated halos reach $10^8 \Msun$ at redshift $z=24$, $10^9 \Msun$ at $z=20$,
$10^{10} \Msun$ at $z=16$ and $10^{11} \Msun$ at $z=10$). Thus we will also
investigate a scenario with higher $z_{vir}$ for the above mentioned range in
halo masses, corresponding to $\sim 4$-$\sigma$ peaks. We take $\lambda=0.03$
as the standard spin parameter for our runs. This value is consistent with the
average spin parameter found in cosmological simulations
\citep[e.g.][]{JangCondell2001}. We have also run our simulations using
$\lambda = 0$ or $\lambda=0.05$, and found no significant differences in our
results \citep[see also][]{Bromm2003}. For the runs we take $N = 10^5 -10^7$
particles and a corresponding spatial resolution of a few tens to a hundred
parsec.

An important parameter for our study is the value of the BH feedback
factor. We will consider two extreme values: $\epsilon_{\rm f}=0$
corresponding to the case without feedback and $\epsilon_{\rm f}=0.05$, the
value that has been used to reproduce the observed normilization of the local
$M_{\rm BH}$-$\sigma$ in galaxy merger simulations \citep{DiMatteo2005}. All
simulations are run to a final redshift of about $z=6$. After the collapse
redshift there is no further infall of material or merging in these models, so
the top-hat is not a realiable tool to investigate evolution of structure
after this time.

\begin{figure}
    \centering
    \epsscale{.95}
    \plotone{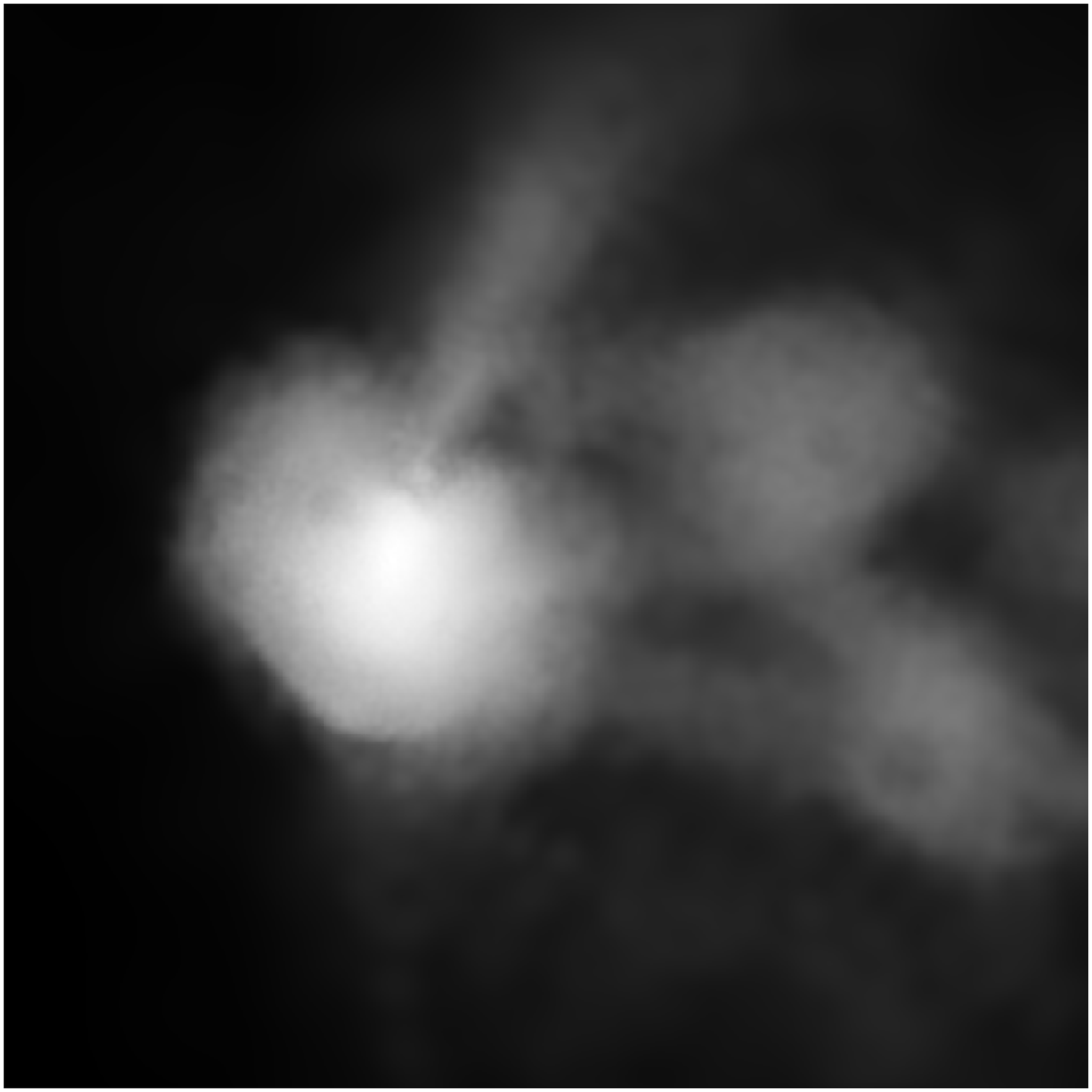}
    \plotone{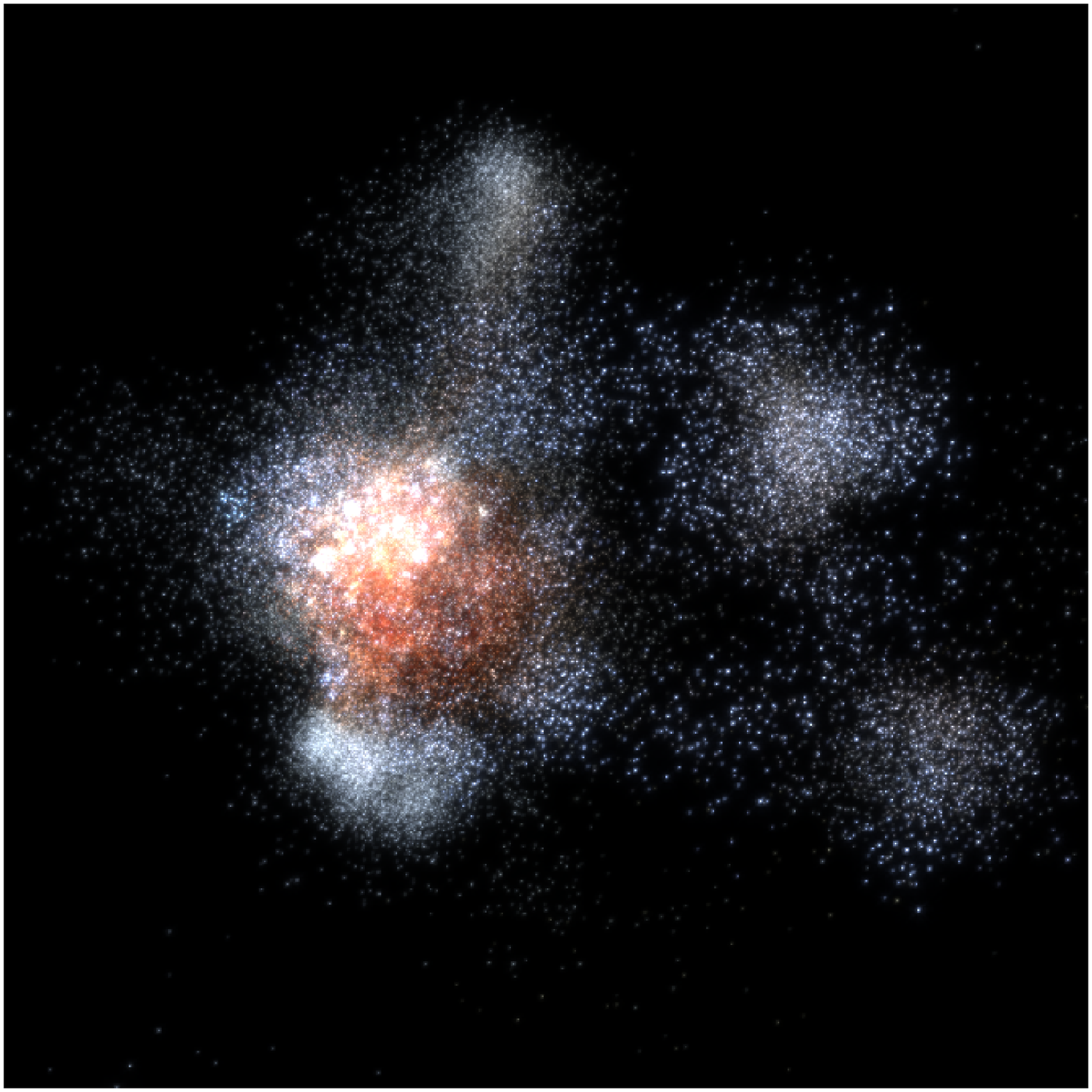}
    \caption{Projection of the gas distribution (top) and rest frame UBV
    composite rendering (bottom). Shown is a region 10 kpc wide of the A3
    run at $z=10$. The UBV panel includes stellar evolution 
    and extinction by the ISM.}
    \label{fig:UBVHI} 
\end{figure}

\begin{figure*}
    \centering
    \epsscale{0.49}
    \plotone{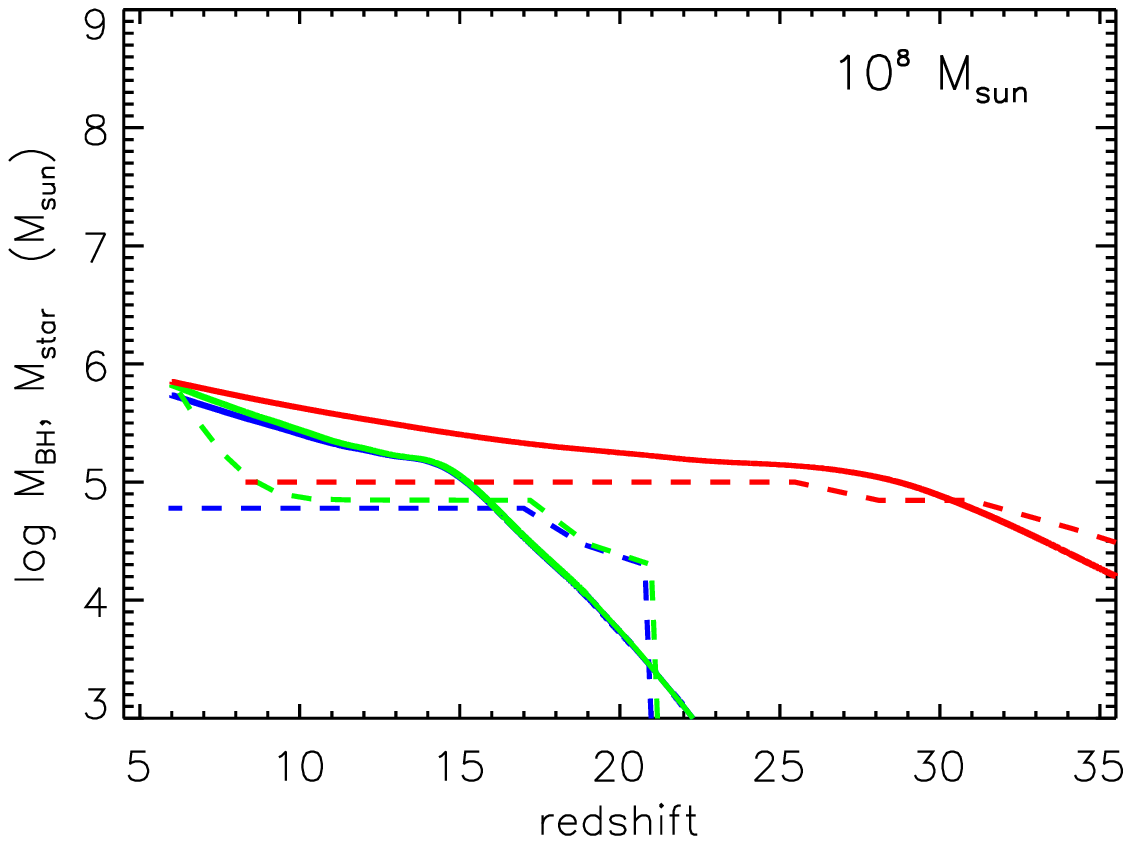}
    \plotone{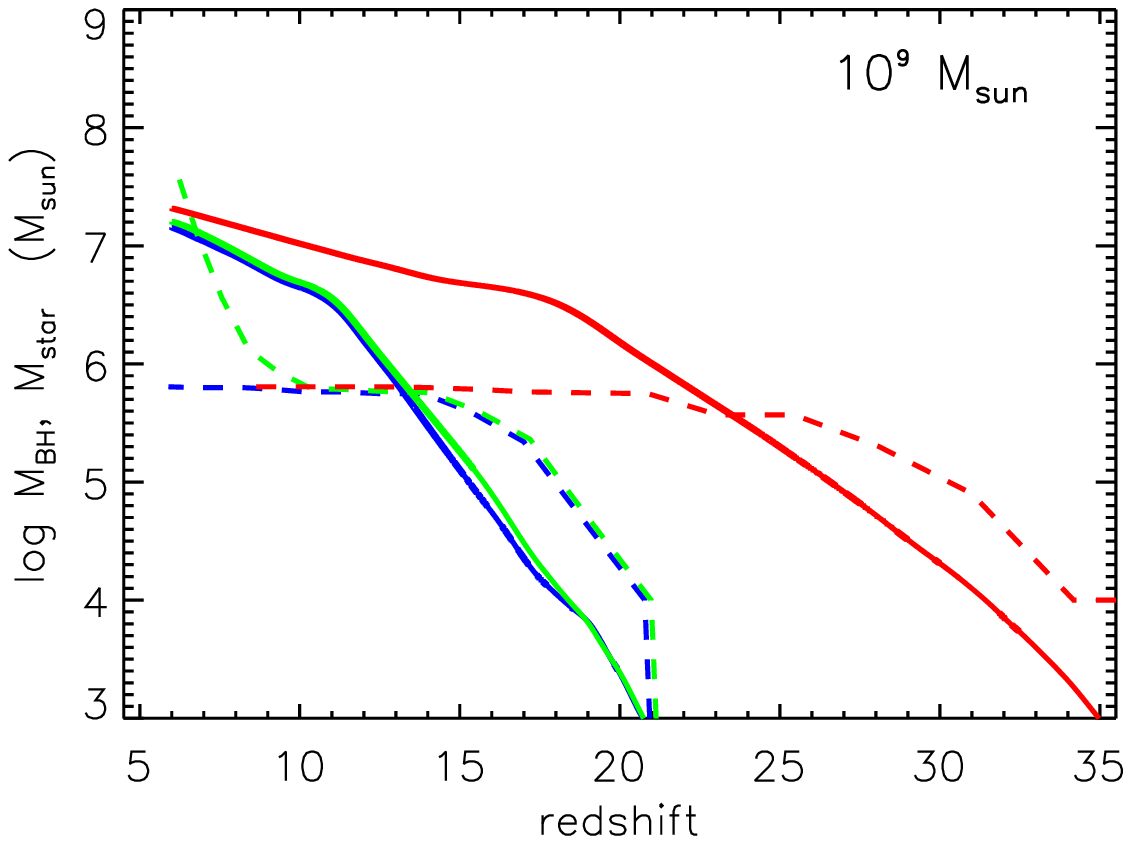}
    \plotone{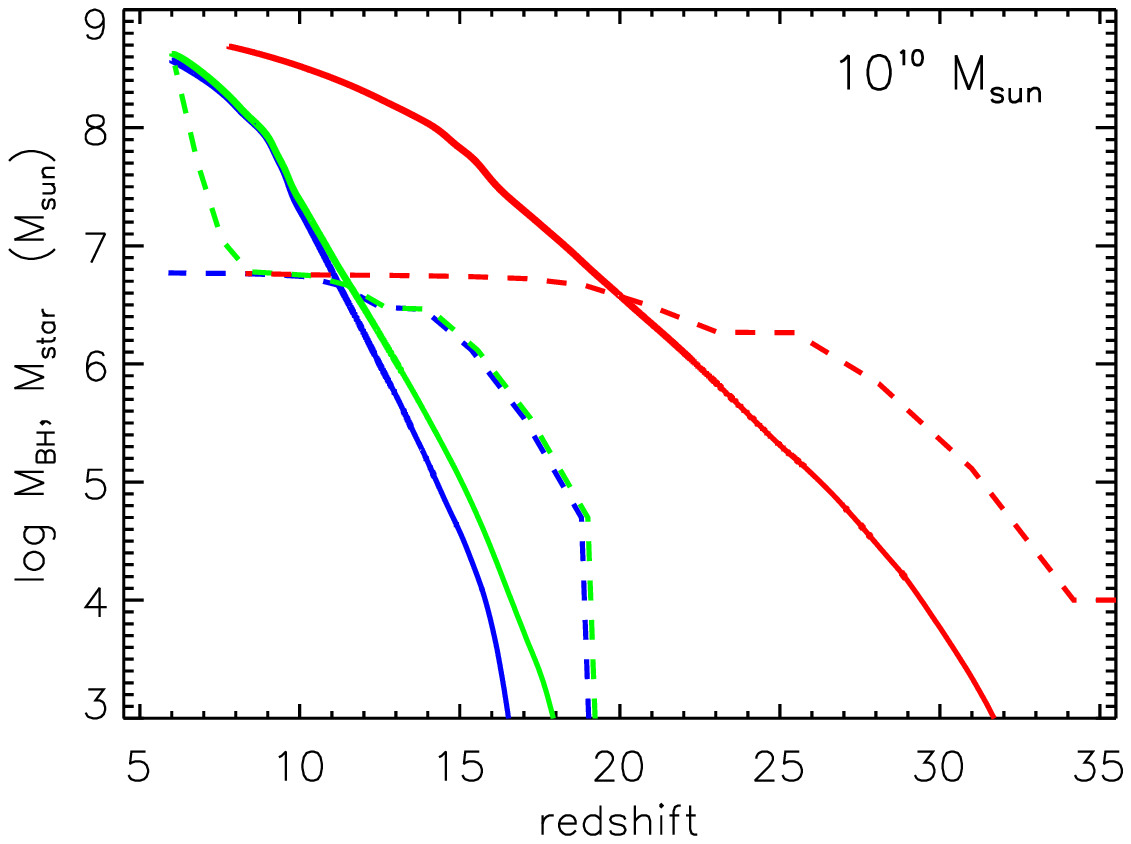}
    \plotone{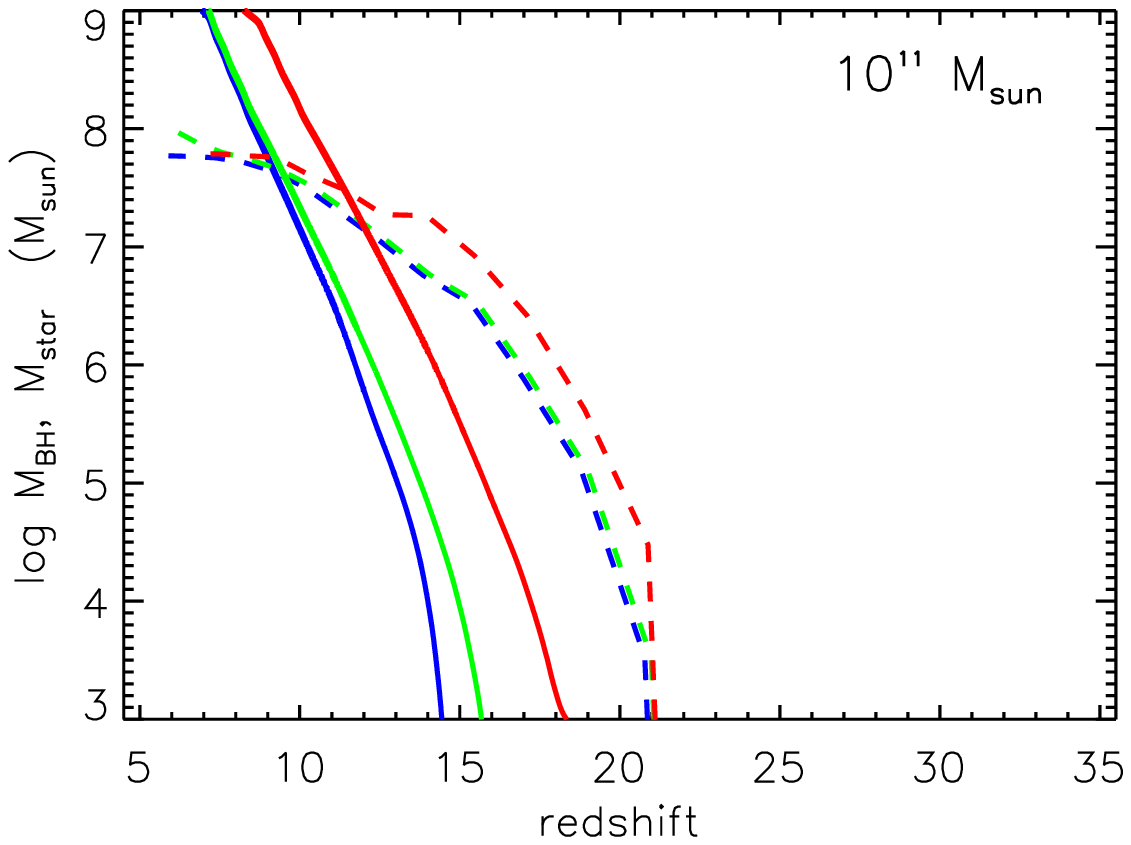}
    \caption{ Evolution of the total stellar mass and black hole mass for the
    main simulations. Different panels show the results for halos of
    different mass (each panel labeled with the corresponding mass), the
    solid lines show the total stellar mass, the dashed lines show the total
    BH mass. Colors indicate the type of simulation: blue line=A (default),
    green line = B (no feedback) and red line = C (higher $z$ collapse).}
    \label{fig:global} 
\end{figure*}

\begin{figure}
    \centering
    \epsscale{1.}
    \plotone{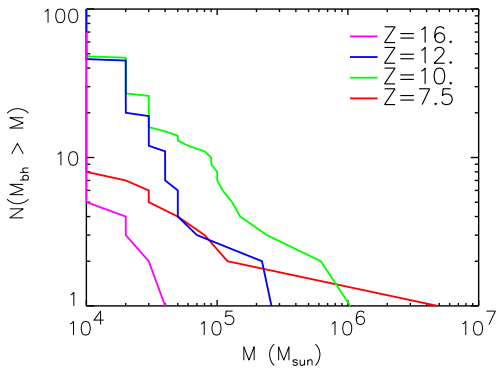}
    \caption{ The evolution of the black hole mass function of the  
     A3 run ($10^{10} \Msun$). The plot shows the number of black holes
     $N(M_{\rm BH} > M)$ per halo of the $10^{10} \Msun$ A3 run as a
     function of mass for four different redshifts in the simulation. }
     \label{fig:massfn}
\end{figure}

\section{Results}
\label{sec:results}

\subsection{The evolution of the central host and the seed black hole population} 
In Figure~\ref{fig:tophat} we show a short sequence of
projections of the baryon distribution (color coded by temperature) of the A3
run, representative of the simulations presented here. The sequence starts at
redshift $z=20$ and ends at the redshift of collapse $z_{\rm vir} \sim
10$. Each panel is 200 kpc (comoving) on the side. As expected, the
over-dense region initially expands with the Hubble flow at a reduced rate and
subsequently turns around (at $z \sim 16-17$) and eventually collapses. The
evolution of the central object, as well as the formation of substructure
follows the general scenario of structure formation in a $\Lambda$CDM
cosmology. Briefly after turn around, the dark matter component has developed
significant substructure and baryons start falling into the deepest
potentials, black holes are formed and star formation eventually ensues (the
only black holes shown in Figure~\ref{fig:tophat} are the most massive and
those twice the initial seed mass). Close to the redshift of virialization,
$z_{vir} \sim 10 $ the halo undergoes violent relaxation, with the gas
settling in the central regions.

Figure \ref{fig:UBVHI} shows a more detailed view (10 kpc proper) of the gas
distribution and a simulated rest-frame UBVI image composite just before
final collapse at $z \sim 10$.  At this point, what looks like a heavily
obscured star bursting system fed by remnant bridges and filaments of HI
gas, is about to merge with two other smaller systems. Blue colors show the
intense star formation going on (the star formation rate is about $1.5
\Msun/{\rm yr}$), indicating the general youth of the system. The total star
mass is about $5 \times 10^7 \Msun$ (see Fig.~\ref{fig:global}). The
remaining massive black holes (that have not yet merged) and their associate
feedback energy give rise to the faint low density bubbles in the gas
distribution (e.g.; in the lower right object of the bottom panel of 
Fig.~\ref{fig:UBVHI}). The total mass of black holes at $z =10$ is $6.3
\times 10^{6} \Msun$, with the heaviest, at the center of the collapsing
halo, accounting for $10\%$ of that. Further evolution of this system will
form a elliptical dwarf galaxy after the gas reservoir is depleted (at $z=6$
the star formation rate has markedly leveled off, see Figure~\ref{fig:global}).

\begin{figure}
    \centering
    \epsscale{1.}
    \plotone{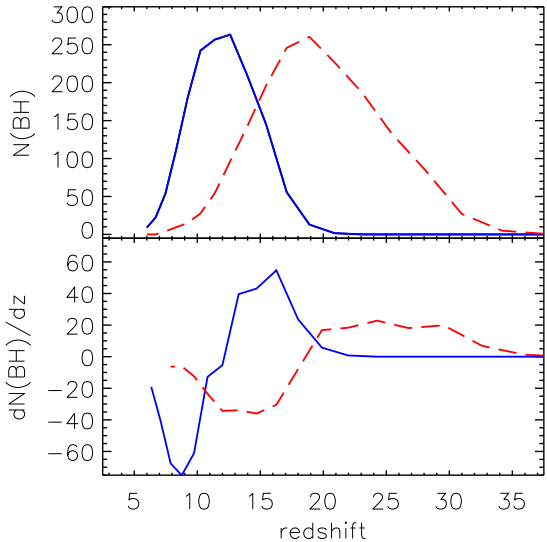}
    \caption{Black hole counts. top panel: total number of BH  
    for the  $10^{10} \Msun$ runs as a function of redshift and bottom panel: 
    BH formation rate. 
    Blue drawn line shows results of the run A3 (normal $z_{\rm vir}$),
    while the red dashed line is for the C3 run (high $z_{\rm vir}$).} 
    \label{fig:bhcounts}
\end{figure}

Within our model the formation of seed BH mirrors the evolution of
substructure in the halo. The growth of the cumulative BH mass function and
the evolution in the number of BHs are shown in Figure~\ref{fig:massfn} and
Figure \ref{fig:bhcounts} respectively. At $z=20$ the number of BH is still
low, commensurate with the lack of structure at masses scales of $10^6 \Msun$
in the parent halo. However as substructure forms in the simulation, the
formation rate of BHs increases, peaking at $z=16$ (as expected this is close
to the turnaround redshift for this halo). Most BHs have just formed and
therefore the mass function peaks at around the seed mass value. At a redshift
$z=12$ (lower left panel of Figure~\ref{fig:tophat}) the growth rate is offset
by the merger rate and the number of BH reaches a maximum. The BH mass
function has grown significantly at most mass scales.

Further evolution of the collapsing halo decreases the number of BHs because
of the merging of sub-halos (and the corresponding mergers of central
BHs). This is reflected by the more significant growth in the high mass tail
of the mass function. Finally, at the redshift of collapse the merger rate of
BH peaks, resulting in a rapidly declining number of BH in the lower mass end
and a fast growth of the central massive object. Also shown in
Figure~\ref{fig:bhcounts} are the resulting BH counts for the C3 run. In this
case the collapse occurs at higher redshift, and the evolution of the number
of BHs happens at correspondingly earlier times. It is otherwise similar to
the A3 case, with the peak formation rate in this case occurring at $z=24$.

As shown in Figure~\ref{fig:global}, up to about the turn-around redshift of
the halo the total black hole mass density is higher than the star density,
but as significant cooling of the gas ensues after this time, star
formation becomes significant and the resulting star mass density quickly
exceeds the black hole mass after $z\sim 12$, the growth of which has
virtually saturated by $z\sim 10$. This is also reproduced in the C3 run,
where the cross over point for the star density to exceed the black hole
mass is at commensurately higher redshift ($z\sim 22$).

\begin{figure}
    \centering
    \epsscale{1.2}
    \plotone{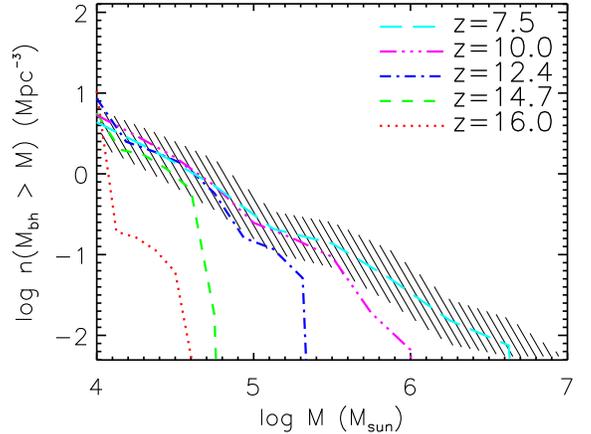}
    \caption{ Global BH mass function at different redshift. Shown is
    the total mass function as derived from our simulations and the
    Press-Schechter formalism, for redshifts from 16 to 7.5. Shaded area
    gives the range of uncertainty from Poisson errors for $z=7.5$. Similar 
    ranges of uncertainty apply for the other redshifts, but are omitted for
    clarity.}
    \label{fig:totalmf}
\end{figure}

 We can obtain an estimate of the global black mass function by taking our
simulations to be representative for halos of that mass and multiplying the
corresponsing number of BHs in each black hole mass bin by the expected number
density of halos of that mass and summing over the mass range of our
simulations,
\begin{equation}
\label{eq:mf}
n(M_{\rm BH}>M)=\sum_i n_i N_i(M_{\rm BH} > M )
\end{equation}
with number densities $n_i$ estimated using the Press-Schechter formalism.
This is shown in Figure \ref{fig:totalmf}. The basic picture that emerges
from this is that low mass BH population builds up early, and as heavier 
halos collapse the range gradually expands to higher masses, in agreement 
with Figure \ref{fig:massfn}. The low end of the mass function does not fall
(or at least not as much as in Fig. \ref{fig:massfn}), because the high mass
halos contribute comparatively less to the low mass population of BH.

\begin{figure*}
    \centering
    \epsscale{1.}
    \plotone{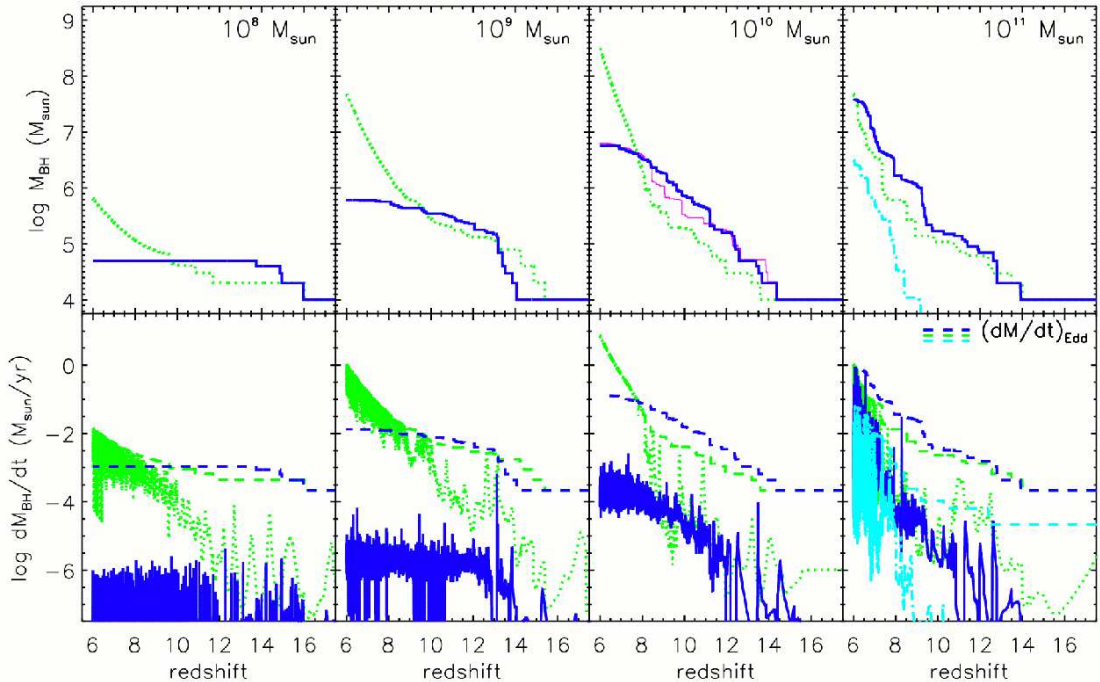}
    \caption{Mass and accretion rate of the central black hole for
    simulations of different halo mass and varying feedback strength.
    The mass of the parent halo is labeled in each panel. 
    The top panels show the total mass as a function of redshift 
    for the most massive black hole. Bottom panels show the
    corresponding accretion rates (drawn lines) as well as the 
    Eddington accretion rates (dashed). In all panels, the blue line 
    indicates simulations with a feedback strength of $\epsilon_{\rm 
    f}=0.05$ (A runs), the green dotted lines those 
    with $\epsilon_{\rm f}=0$ (B runs). An intermediate case for
    $\epsilon_{\rm f}=0.01$ is also shown (purple thin line) for the
    $10^{10} \Msun$ simulation.  The dash-dotted line in the 
    $10^{11} \Msun$ panel shows results for a reduced seed mass 
    ($M_{\rm seed}=10^3 \Msun$).}
    \label{fig:mbhvstime}
\end{figure*}

In the next section we will discuss our results on the growth history of the
massive black hole in the center of the collapsing halo as a function of halo
mass.

\begin{figure}
    \centering
    \epsscale{1.}
    \plotone{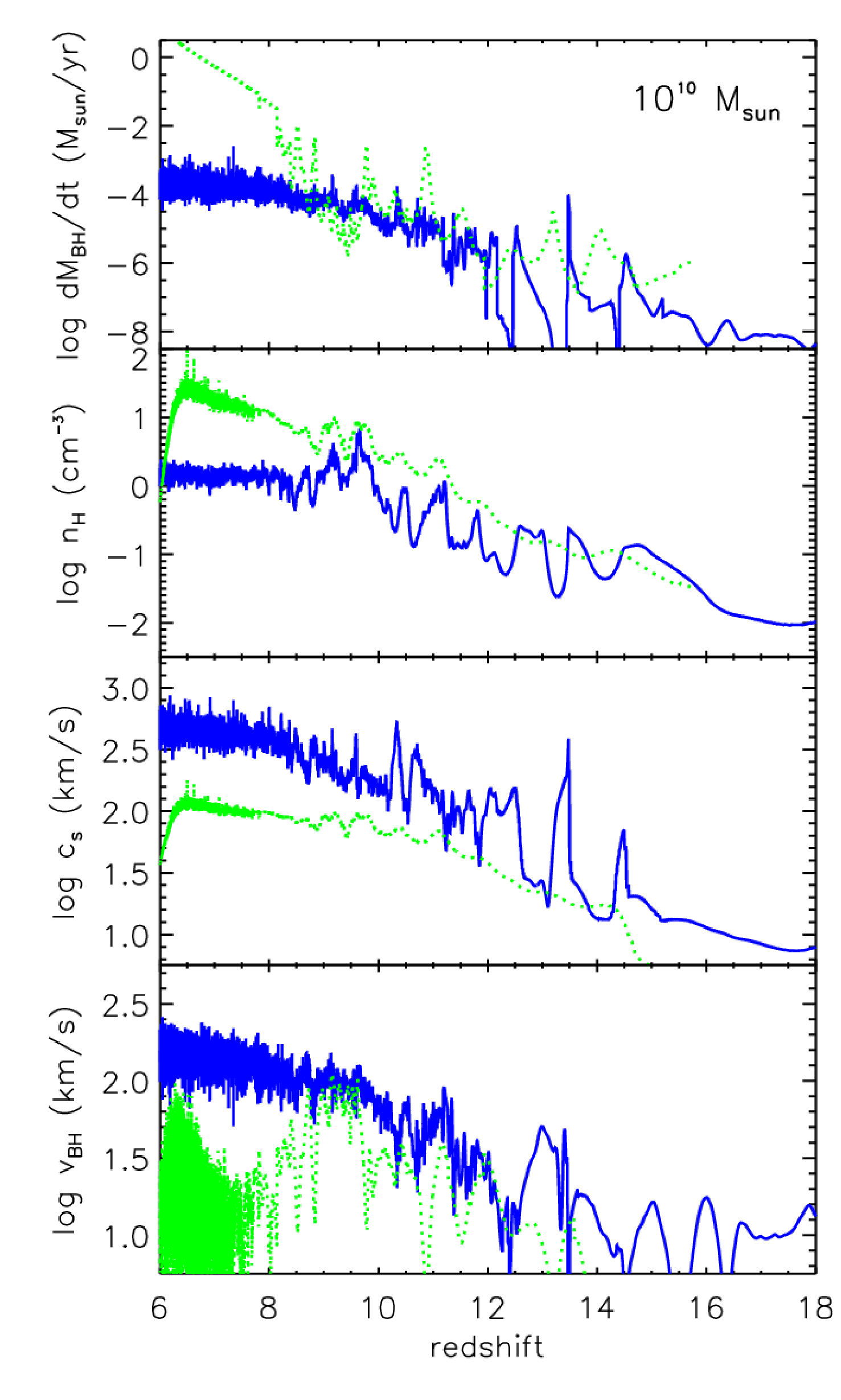}
    \caption{Evolution of the ISM properties around the BH.
    Plotted are (from top to bottom)
    BH accretion rate, local gas density, sound speed and the relative
    velocity of the BH with respect to the gas for the A3 (blue lines) and 
    B3 (green dotted lines) runs.}
    \label{fig:bhism} 
\end{figure}

\subsection{BH growth and accretion history}

Figure \ref{fig:mbhvstime} shows the evolution of the black hole mass (top
panels) and accretion rate (bottom panels) for the most massive black hole
formed at the center of the collapsed halos. The major BH is traced through
its most massive progenitor back to the first seed, which forms at a redshift
of around $z \sim 20$ for the majority of our simulations.  The accretion rate
and the black hole mass are shown for runs for which we have included BH
feedback and for runs with with no BH feedback (A1-A4 and B1-B4 in
Table~\ref{tab:runs}). For the $10^{10} \Msun$ halo, the results for an
intermediate feedback strength ($\epsilon_{\rm f}=0.01$) are also shown.

If we compare the runs of different mass halos with and without feedback the
impact of BH feedback becomes clear: even with modest levels of feedback
($5\%$ of the radiative energy in most cases,  also shown is the  case
of $1\%$ for the $10^{10} \Msun$ halo) the growth of the black hole  is
severely limited compared to the case with zero feedback.
In the models with no feedback the accretion rate starts off at values of the
order or $10^{-6} \Msunpyr$,  much the same as the models with
feedback. However, while in the feedback runs the average accretion rate stays
well below the critical values (by at least a factor of 100), in the  zero
feedback case the typical accretion rates reach the Eddington values leading
to a fast growth of the black hole mass. Even in this case, however, the
critical accretion phase is reached only close and subsequent to the time of
final collapse of the halos (i.e. only at $z < 10$); accretion is
sub-Eddington before that. As shown by the discrete jumps in the black hole
mass evolution in Figure~\ref{fig:mbhvstime}, the growth of the central black
hole is dominated by black hole mergers up to the time it reaches the critical
accretion phase. In models with BH feedback, gas accretion also becomes
increasingly more important as the halo mass increases (and hence the central
gas densities increase).   As the accretion rate increases and the BH
becomes heavy enough, the feedback energy associated with accretion, is able
to heat the gas sufficiently to ``shut off' any further growth. This occurs
rather abruptly. The greater the mass of the halo, the longer the simulation
with feedback follows (approximately) the growth curve of the simulation
without feedback. Hence the difference is smallest for the $10^{11} \Msun$
run.  Figure~\ref{fig:mbhvstime} shows that the difference in evolution of
black hole accretion rate and hence final black hole masses in models with and
without BH feedback increases for smaller halo mass runs. For the $10^{11}
\Msun$ run the accretion rates are similar up until the time close to the
collapse (at $z\sim 7.5$). At this point, the growth barely reaches the
critical Eddington value, while lower mass runs with feedback do not reach
Eddington growth at all. So this mass may be considered the minimum mass for
any efficient black hole growth relevant to quasar activity.  Still, only a
fraction of the final black hole mass ($\sim 10\%$) results from mass accretion
 in this case. At lower mass scales the growth is more inefficient still,
occurring mostly due to BH-mergers (for the $10^8 \Msun$ case less than $1\%$
is due to accretion). This means that the masses derived are sensitively
dependend on the assumed seed masses.

Note that for the runs without feedback the BH starts to grow at the Eddington
rate eventually, not from the onset of BH formation, but only after a
significant number of mergers have occurred. When growth at the Eddington rate
does start, an exponential runaway growth phase occurs. In this case $\lesssim
5\%$ of the final mass of the BH consists of the original seed masses.  At the
end of the simulations, we see that the final BH masses are proportionally
larger for larger halo masses, except for the $10^{11} \Msun$ simulation,
because in this case the time after the $z_{\rm vir}$ is comparatively much
shorter than for the other simulations, leaving too little time for all the BH
mergers and much of the significant accretion phase to occur (as we have
stopped all simulations at $z_{end} =6$).

In order to better appreciate the cause of the differences in the BH accretion
history for the cases with and without feedback, we show in
Figure~\ref{fig:bhism} the evolution of the physical quantities that determine
the growth of the black hole in our model. The gas density around the BH in
the model with feedback has lower density, as well as higher sound speed than
the in run without feedback. This is expected due to the local heating
associated with the BH feedback which eventually energizes the gas
sufficiently to drive a slow wind
\citep{Springel2005b}. Figure~\ref{fig:bhism} also shows that the relative
BH-gas velocities are higher in the feedback case: if no feedback is included
the gas settles in a quiet equilibrium which is most efficient at feeding the
central BH (note that the decrease in density and sound speed at the end of
the B1 run is a result of gas depletion due to runaway BH growth). The effect
of BH feedback on its surroundings is also visible in the much stronger
variations in density and sound speed, indicating that the ISM around it is
being stirred, and definitely not in equilibrium. This is consistent with the
gas density projection in Figure~\ref{fig:UBVHI} which shows the faint bubbles
blown in the ISM as a result of the BH feedback.

\begin{figure}
    \centering
    \epsscale{1.2}
    \plotone{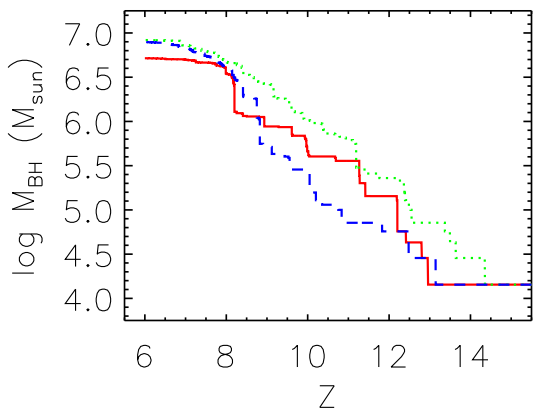}
    \caption{Resolution test for the $10^{10} \Msun$ A3 run. Plotted is the 
    mass of the most massive black hole in the simulation, for runs of 
    different resolution, blue dashed line: (standard) $N=10^6$, red drawn
    line: $N=10^5$ and green dotted: $N=5 \times 10^6$.}
    \label{fig:restest}
\end{figure}

\subsubsection{Resolution tests}
\label{sec:restest}

In order to assess possible resolution effects on our results, we have carried
out resolution tests. Figure~\ref{fig:restest} shows the black hole mass for
the $10^{10} \Msun$ halo with three different particle numbers
($N=10^5$,$10^6$, $5\times 10^6$) spanning a factor of 50. The results show
consistent black hole growth. In particular, the final BH mass is rather
insensitive to the resolution used. The lowest resolution run ends up with
40\% lower BH mass due to the fact that there is some loss in resolution of
the substructure that would form extra BH at high resolution, but the two
higher resolution runs show convergence for the final BH mass within
10\%. Even though at high redshifts there is some disparity between the runs
at different particle numbers, there is no systematic trend with particle
number. Some differences are expected due to the stochastic nature of the BH
merger growth in our simulations, which tend to magnify small differences in
BH trajectories at different resolution. We consider our results to be
reliable at the medium resolution, although for highest mass halos we use the
higher particle count.

\section{Discussion}
\label{sec:disc}

\begin{figure}
    \centering
    \epsscale{1.2}
    \plotone{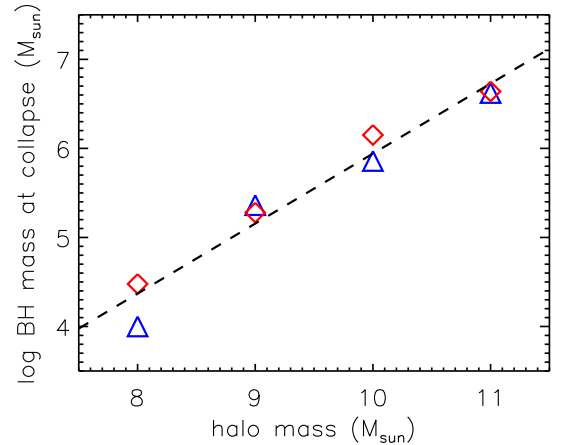}
    \caption{BH mass and halo mass. Plotted is the mass of the central most
    massive BH at the redshift of virialization of the parent halo as a 
    function of halo mass. Blue triangles are the simulations of the ``A'' 
    series (our default 3-$\sigma$ runs), while the red diamonds are the 
    ``C'' simulations (high collapse redshift). The dotted line is the
    fitted relation $M_{BH} = 5\times 10^6 \Msun
    (M_{halo}/10^{11}\Msun)^{0.78}$. }
    \label{fig:bhvshalo} 
\end{figure}

Our results caution against the commonly made assumption for the early black
hole growth, namely that they accrete at the Eddington rate
\citep[e.g.][]{Haiman2004, Yoo2004, Madau2004, Volonteri2006}. We have shown that this is
likely an overestimate of the growth rate, and may consequently lead to overly
optimistic mass estimates for the central black holes masses of the first
quasars.  In our simulations, seed black holes typically accrete well below
the critical Eddington rates. For halos $M \le 10^{10} \Msun$ BHs never reach
an Eddington growth phase, or if they do, it is only at the time close to the
collapse redshift and under the rather extreme assumption that any kind of
black hole feedback is switched off. The growth reaches the Eddington rate at
late times only for halos of $M \ge 10^{11} \Msun$. The final black hole
masses at the time of collapse are proportional to the total halo mass with
\begin{equation}
\label{eq_bhvshm}
M_{BH} = 5\times 10^6 \Msun (M_{halo}/10^{11}\Msun)^{0.78}
\end{equation}
(Fig.~\ref{fig:bhvshalo}). For the highest mass halo this implies a black
hole mass of $M_{BH} \sim 10^{7} \Msun$, a couple of orders of magnitude
below the values deduced for the $z=6$ Sloan quasars. Our results therefore
imply that the first black holes, remnant from the first generation of
stars, are unlikely to grow into the first quasars  unless the Sloan quasars
are  hosted  in halos rarer than 3-4 $\sigma$ peaks.
If we compare the number density in halos inferred from the space density of
quasars at $z =6$ (Fan et al. 2004) this requires that their host masses be
larger than those studied here, and of the order of $10^{13} \Msun$ typical of
an extremely rare 5.5-$\sigma$ (or higher for WMAP3) peak (note however that
this assumption corresponds to a duty cycle of order unity for the Sloan
quasar). An extrapolation of Figure~\ref{fig:bhvshalo} (using Eq.
\ref{eq_bhvshm}) gives a BH mass of $\sim 10^8 \Msun$ for a $10^{13} \Msun$
halo (independent of $z_{\rm vir}$), which falls short by a factor of $\gtrsim
10$ with respect to the inferred SDSS quasar masses. If taken at face value,
Equation~\ref{eq_bhvshm} implies a minimum halo mass of $10^{14} \Msun$, which
represents a $>8$-$\sigma$ peak at $z=6$. Such peaks are much too rare. An
earlier collapse redshift for the $10^{13}~\Msun$ halo may allow the BH mass
to grow to the SDSS values (assuming the Eddington growth is sustained after
the virialization of the parent halo). The collapse would then be required to
occur at least $\Delta z \approx 0.5$ earlier. This shifts the $5.5$-$\sigma$
peak to even higher $\sigma$ and again makes the parent halo a factor $20-50$
too rare.  

These rather stringent conclusions on the inefficiency of seed growth are
reached despite the fact that we have made rather optimistic assumptions
regarding BH growth. Namely, we have assumed that all halos of a certain
mass will host a seed BH (thus without taking into account possible feedback
effects from the formation events of these BHs on nearby halos).
Furthermore, the seed BHs we assume to be formed are massive, certainly in
the upper range of masses considered for BH seeds from remnant Pop III
stars. However, the rather slow black hole growth implied by our results
makes the problem a rather strong function of the initial choice of seed
mass (as most of the growth occurs at the Bondi rate or via mergers instead
of an exponential Eddington phase).   Smaller seed masses than the value
chosen here would  imply even slower growth. This is illustrated in Figure
\ref{fig:mbhvstime}, where , for the $10^{11} \Msun$ run we have also
plotted a rerun of the A4 simulation,  but with a seed mass of $M_{\rm
seed}=10^3 \Msun$. In this case the final mass is more than a
factor 10 smaller, as expected. Note however that the BH in this case also
reaches Eddington growth. This again shows that a halo mass of $\sim 10^{11}
\Msun$ is sufficient for Eddington growth. Larger seed masses would allow
faster growth. However, larger seeds (e.g. $\sim 10^5 \Msun$) than the ones
we have chosen would merely emphasize the need of a jump-start scenario for
the formation of the $z=6$ Sloan quasars.  Using detailed simulations,
\citet{Li2006} have shown that halos of $10^{13} \Msun$ could be the hosts
of the first quasar. However, in their work it was assumed that  BH
would undergo growth at the Eddington rate, preceding the time of major
mergers at $z \sim 7-12$.  This isolated critical growth phase implies
masses $\sim 10^5 \Msun$ upon  entering major mergers.  Furthermore they
used fully assembled galaxies models as their initial conditions, thus not
following the assembly of galaxies and their BH self-consistently.  

One feature of our modeling that stands out as being inconducive to BH growth
is the inclusion of BH feedback. Even though the exact mechanism by which the
AGN acts on its surrounding ISM is poorly understood, it is likely that the
end-effect of AGN feedback is that a fraction of the accreted rest mass energy
is thermalized in the surrounding medium, which is what we consider. The
amount of feedback energy we use is a rather moderate $0.5\%$ of the rest mass
of the infalling material. We have also shown that simulations with $5\times$
lower feedback strengths (see Fig.~\ref{fig:mbhvstime}) yield qualitatively
similar results. Any amount of black hole feedback is therefore expected to
affect the black hole growth rather dramatically regardless of the details of
the feedback processes.

The multiphase model we use for the star forming ISM takes into account
that we have limited resolution to follow the densest phases of the ISM.
Nevertheless, this introduces some dependence on the choice of model
parameters. For example the critical density for star formation might be a
function of the chemical composition. Higher critical densities would give
a higher mean density for the ISM, leading more rapid growth. It is
unlikely, however, that the star formation model in these early objects is
radically different, because star formation is locally seen to proceed in much
the same way for a wide variety of environments, amongst which very low
metallicity ones. 

For our choice of parameters almost all BH mass ends up in one central BH
(Fig. \ref{fig:bhcounts}). This is, to some extent, the result of our
simplified criterion that leads to BH mergers. Clearly, different assumptions
are likely to decrease the merger rate, even though, in general, the view is
that the friction processes that bring two BHs into close vicinity of each
other will operate efficiently
\citep[e.g.][]{Begelman1980,Chatterjee2003}. Here we do not take into account
the effects of momentum carried away by gravitational waves. This so called
``gravitational recoil'' (GR) can expel BH from their parent halos instead of
merging them. This effect was examined in semi-analytical models by
\cite{Haiman2004}, \cite{Yoo2004} and more recently by
\cite{Volonteri2006}. Haiman found that GR will inhibit BH growth unless
recoil velocities are smaller than $\sim 65 {\rm km/s}$. \cite{Yoo2004}, on
the other hand, managed to grow BH to $10^9 \Msun$ by $z\approx 10$ for
slightly different assumptions regarding the GR strength and the escape
velocities from halos, with about equal contributions from BH mergers and gas
accretion processes. In either case, the authors did however make the
assumption that the fueling of the BH is sufficiently effective to support BH
growth at the Eddington rate. Interestingly, \cite{Yoo2004} have examined the
effects of restricting the gas accretion to halos of a minimum velocity
dispersion.  They found that for the upper ranges of the threshold 
dispersion they considered the BH growth was severely impeded, 
which is consistent with our findings. Such a threshold would be somewhat
ad-hoc in their model, whereas we find that indeed this restricted mode of
 accretion is more realistic due to the feedback effects. 

\begin{figure*}
    \centering
    \epsscale{0.49}
    \plotone{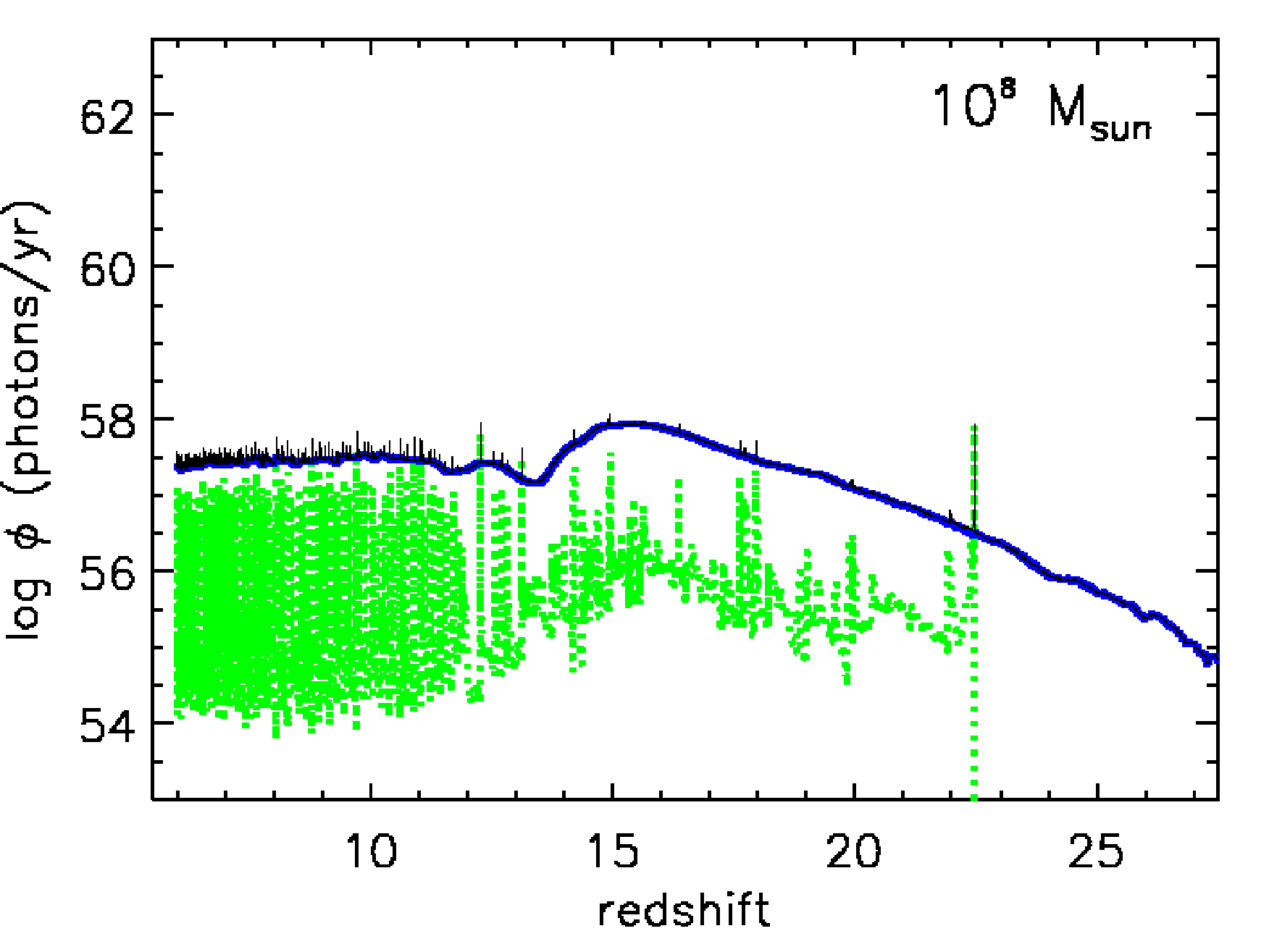}
    \plotone{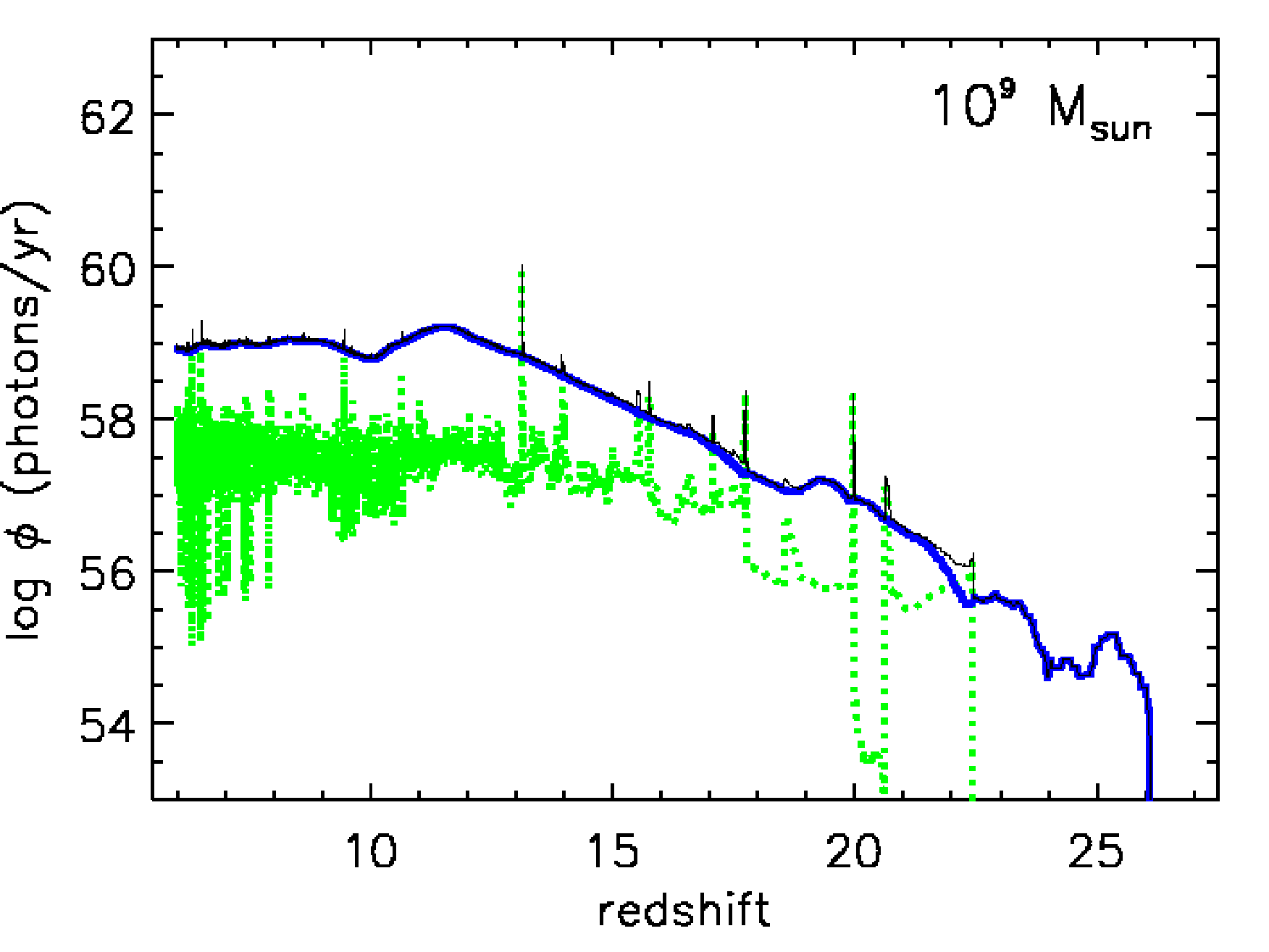}
    \plotone{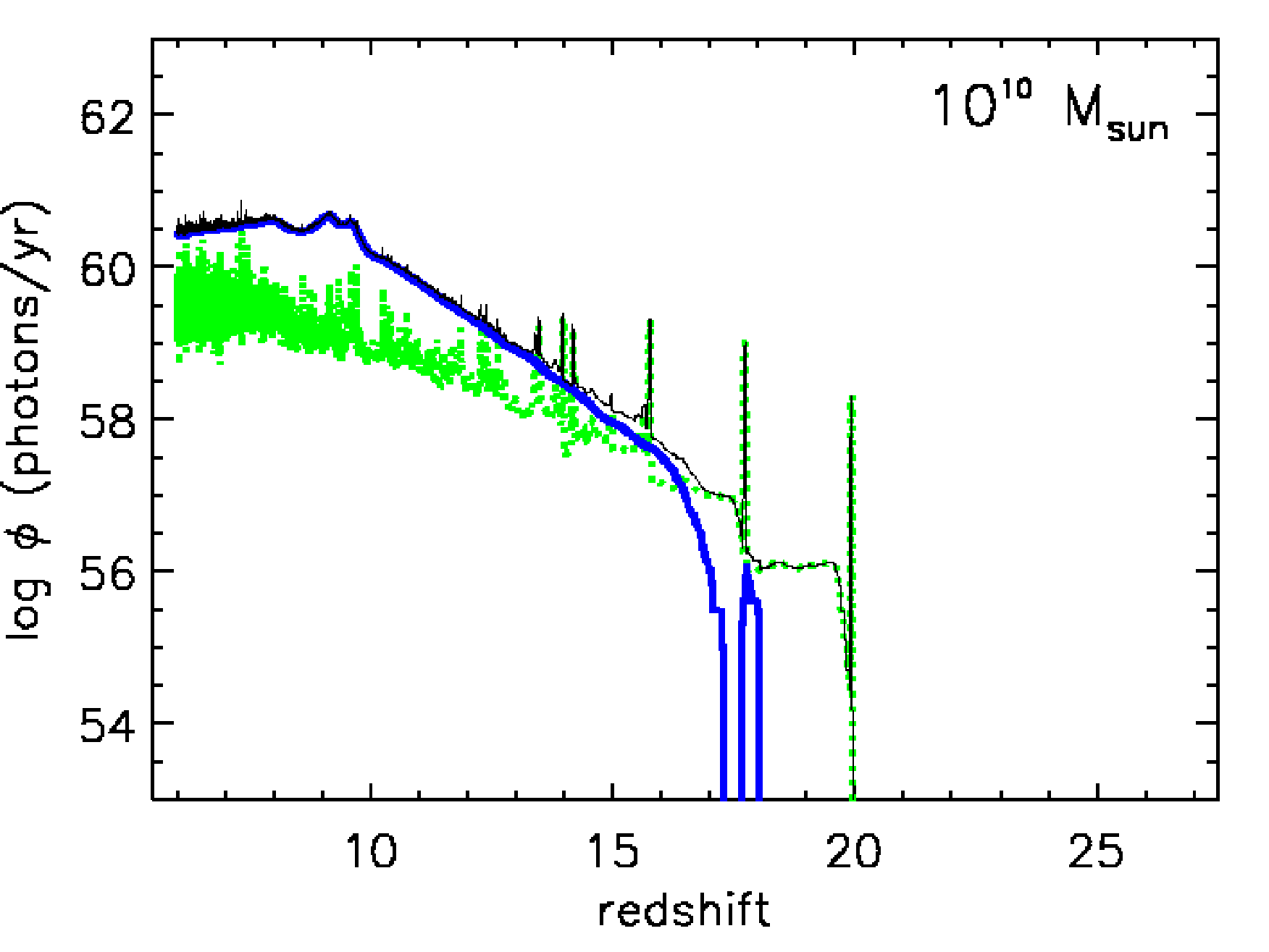}
    \plotone{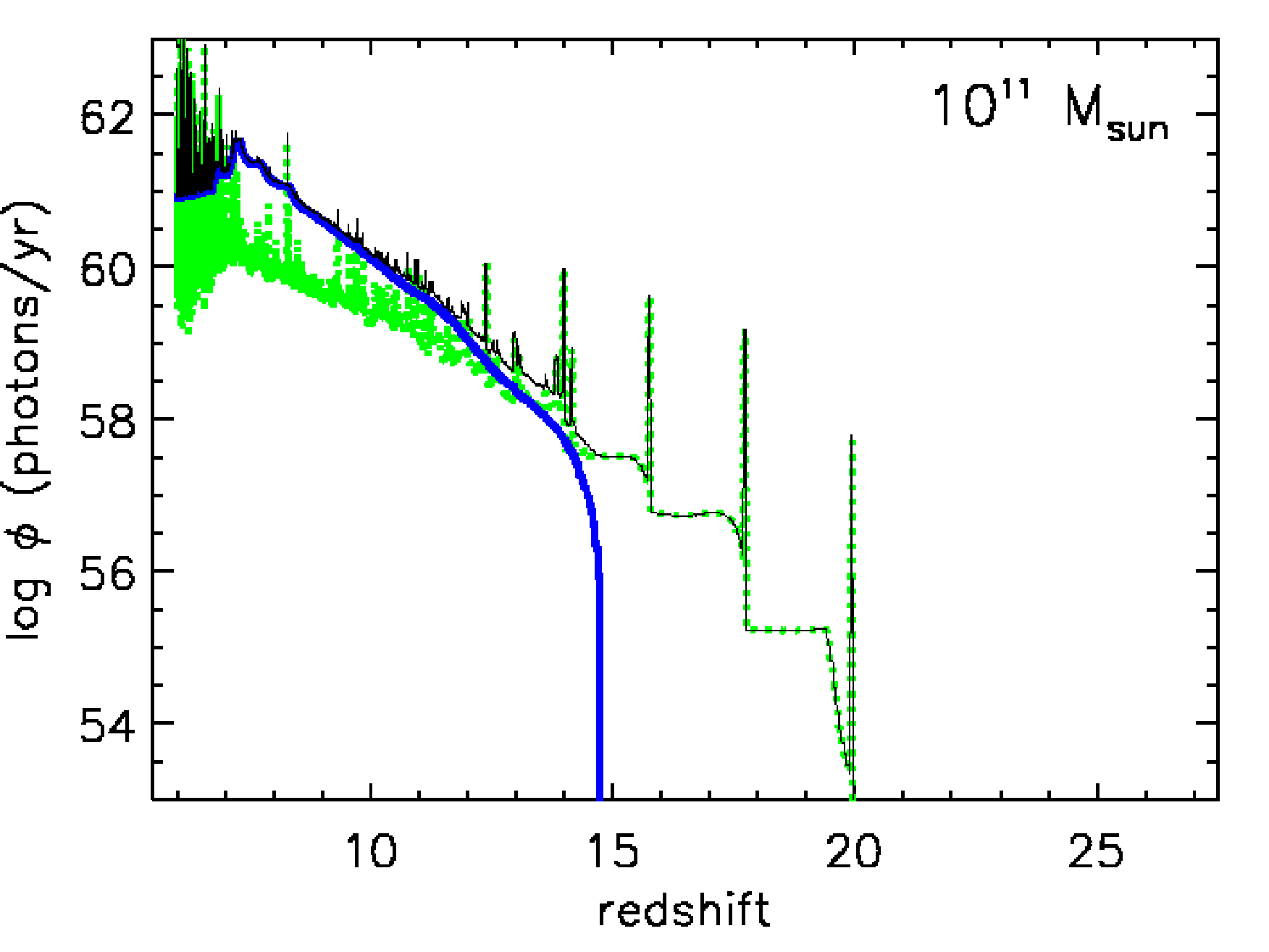}
    \caption{Ionizing photon production. Plotted are in blue drawn lines: 
    the production rates $\phi_{\star}$ of ionizing photons for the stars,
    in green dotted: $\phi_{\rm BH}$ for the ionizing flux from all 
    black holes and thin black drawn line: the total 
    $\phi=\phi_{\star}+\phi_{\rm BH}$. For the stellar 
    ionized photons it is assumed that photons are liberated 
    instantaneously at the formation redshift.}
    \label{fig:ion} 
\end{figure*}

\citet{Volonteri2006} explore the conditions for the growth of seeds up to the
SDSS quasars again assuming that this occurs at the Eddington rate. In
particular, they consider the impact of dynamical effects of gravitational
recoil. Their main conclusion is that the occurence of SMBH at early times 
is very selective. They argue that SDSS quasars could be hosted in halos of 
$M \sim 10^{11} \Msun$, provided that the radiative efficiency is not too high,
as in this case they are likely not to suffer significantly from the negative 
contribution due to dynamical effects. 
Alternatively for growth in higher halo masses ($10^{13} \Msun$ or so) the 
kick velocities need to be considered at the low end.  Our results however 
highlight the difficulty of sustaining Eddinton growth in the first place in 
halos of $M \sim 10^{11} \Msun$, arguing for even more selective conditions 
that allow for rapid growth.

Faced with the difficulty of growing seed BHs to the Sloan quasars, some
 authors propose that short phases of super-Eddington growth can jump start
 the growth of BHs. In particular, \cite{Volonteri2005b} proposed a scenario
 in which BH grow in thick and dense ($n=10^4\, {\rm cm}^{-3}$, $T=8000$ K)
 gas disks formed at very high redshift, at $z\sim 20$ before the universe is
 enriched and metal cooling can affect collapse and fragmentation. These
 authors argue that the BH should accrete gas at rates comparable to the Bondi
 estimate, which for these conditions far exceeds the Eddington rate. In our
 model, the implicit assumption for the ISM is that some enrichment has taken
 place as a result of the PopIII star formation, significantly enough that
 star formation is proceeding and feedback from SN and stellar winds
 pressurizes the ISM, inhibiting the formation of large and dense central
 disks as those inferred by \cite{Volonteri2005b}. We cannot therefore
 directly probe the scenario of super-critical accretion proposed but
 ultimately the conclusion from that work also requires extremely rare host
 for the Sloan quasars.

In our simulations we have not treated the detailed physics of HII regions
of the first stars in the ISM surrounding the first black holes at
$z\sim 20$ (rather we have  optimistically assumed efficient HI cooling).
\citet{Johnson2006} have recently shown that the seed black hole growth
may experience an additional early bottleneck due to significant delay
between black hole formation and the onset of efficient accretion in the
primordial HII gas. This work strengthens our results, and in general the
argument that it may be hard to grow the first quasars from the remnants
PopIII black holes.

\subsection{Reionization and X-ray heating}
Several authors have recently investigated the potential impact of a
primordial generation of BHs on the reionization process and the thermal
evolution of the IGM. More specifically, high-$z$ BHs could produce a
substantial amount of ionizing photons due to their potentially
 higher luminosity compared to stellar type sources (per unit mass accreted) 
and likely higher escape fraction \citep[e.g.][]{Madau1999, Valageas1999, 
Miralda2000, Wyithe2003c, Madau2004}. 
In addition, their harder spectra would also induce reionization of HeII,
otherwise unattainable with stellar type sources (with the exception of very
massive PopIII stars). Finally, X-ray photons could produce a fairly
homogeneous reionization (due to their large mean free path), albeit not
complete \citep[e.g.][]{Oh2001, Venkatesan2001, Madau2004, Ricotti2004}, and
substantially heat the IGM. Most of these calculations however have also been
carried out under the assumption that the first black holes are growing at the
critical Eddington rate at early time. Here, we briefly investigate the
implication of the black hole accretion history we obtain from our simulations
for ionization and X-ray heating of the IGM. We will limit ourselves to some
simple estimates, deferring a more detailed investigation to a future paper.

The production rates of ionizing photons $h\nu>13.6$~eV is plotted in
Figure~\ref{fig:ion} for the various halo masses we have simulated. The rate
of ionizing photons from stars $\phi_\star$ is calculated as
$\phi_\star=\psi_\star \dot{M}_\star$, where $\dot{M}_\star$ is the star
formation rate and $\psi_\star=3 \times 10^{60}$~photons$/\Msun$ is the yield,
typical of a stellar population with salpeter IMF and a metallicity of about
$1\%$ Solar. In the case of metal-free, massive stars, $\psi_\star$ can be
more than an order of magnitude higher (as discussed above our assumption is
that in general some enrichment has taken place). Thus, the stellar
production rate of ionizing photons should be regarded as a lower limit. We
calculate the BH photon production rate assuming that all radiation comes out
as ionizing photons with a mean energy of $\langle h\nu \rangle = 80$~eV (such
a hard spectrum is justified for smaller mass BH, e.g.
\citet[][]{Madau2004}). This gives a production rate for BH of $\phi_{\rm
BH}= \psi_{\rm BH} \dot{M}_{\rm BH}$, with $\psi_{\rm BH}=1.5 \times
10^{63}$~photons$/\Msun$.

From Figure~\ref{fig:ion} it is clear that the stellar contribution to the
overall budget of ionizing photons is dominant, with the exception of
$z\sim17-20$ in the larger objects. Note some episodes of high luminosity
occur at high redshift for the BH, when accretion rates peak momentarily
before being quenched by BH feedback (such episodes of increased accretion
are also visible in Fig.~\ref{fig:mbhvstime}), but they are too short too
affect the ionization much. It should be noted, however, that the escape
fraction (i.e. the fraction of emitted ionizing photons that are able to
escape in the IGM) for stellar photons is likely to be smaller than that for
photons produced by BH, which might thus be relevant during certain epochs of
structure evolution.

\begin{figure}
    \centering
    \epsscale{1.}
    \plotone{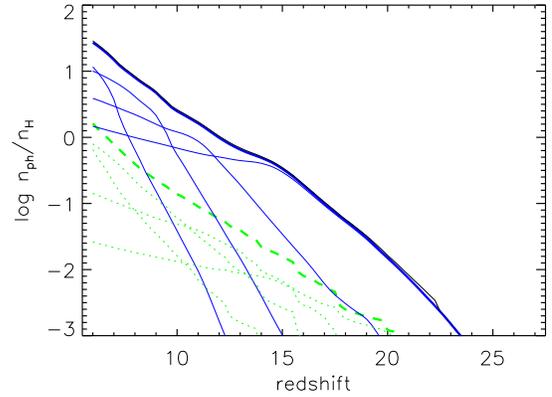}
    \caption{Cumulative number of ionizing photons per hydrogen atom. 
    Plotted as a function of redshift in the blue drawn line: total number of
    ionized photons per hydrogen atom produced by stars  $n_{\phi, \star}/n_{\rm H}$,
    in green dashed: $n_{\phi, {\rm BH}}/n_{\rm H}$ and thin black 
    drawn line: $n_{ph}/n_{\rm H}=(n_{\phi, \star}+n_{\phi,{\rm BH}})/n_{\rm H}$.
    Drawn and dotted lines give the `contribution' of the halos (A1-A4) of
    different masses towards the averaging procedure. }
    \label{fig:nphnh} 
\end{figure}

We estimate the total production of ionizing photons $n_{ph}$, 
similar to our estimate of the total mass function,  by
multiplying the cumulative production of ionizing radiation of each halo by
their expected number density and summing over the mass range of our
simulations, 
\begin{equation}
\label{eq:nph}
n_{ph}(t)=\sum_i n_i \int_0^t \phi_i dt 
\end{equation}
with number densities $n_i$ estimated using the Press-Schechter formalism. 
Figure~\ref{fig:nphnh} shows the contribution of stars and BH to the
cumulative number density of ionizing photons (normalized on the global
hydrogen number density). The contribution from the stars under the above
assumption for the ionizing photon flux and with an escape fraction of 100\%
for both components is an order of magnitude larger at all redshifts. If
the gas is distributed uniformely and does not recombine, we have a
reionization redshift of $z \sim 12$. This conclusion is not affected by
the contribution to ionization from the BHs.

\begin{figure}
    \centering
    \epsscale{1.}
    \plotone{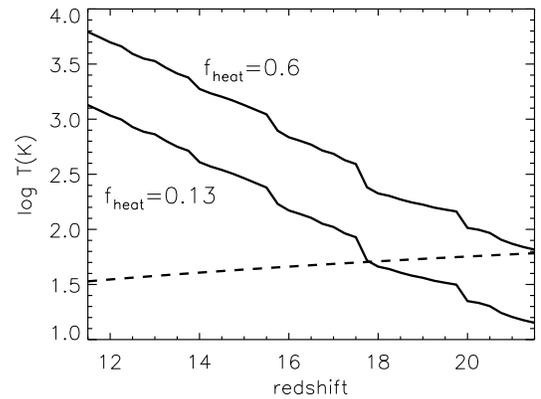}
    \caption{Temperature due to X-ray heating. Drawn lines give the
    temperatures of the IGM if heated by the X-rays generated from BH
    accretion (Eq.~\ref{eq:xraytemp}) as a function of redshift for two
    different values of the heating fraction, $f_{\rm heat}=0.13$
    (appropiate for a low ionization fraction $x=10^{-4}$) and $f_{\rm
    heat}=0.6$ ($x=0.1$). Dashed line gives the CMB background temperature.}
    \label{fig:xheat} 
\end{figure}

One of the most interesting effects of the harder radiation by BHs is IGM
heating due to X-rays. As X-rays are much more likely to be absorbed by HeI
rather than HI (due to the higher cross section), direct photoionization of HI
by X-rays is negligible. Nevertheless, the electron ejected by HeI
photoionization is more likely to ionize HI than HeI because the former is
much more abundant. The partitioning of a photoelectron energy between heating
the gas and secondary ionization is a function of the gas ionization
fraction. For values above 10\% most of the energy goes into heating. For this
reason, although HI ionization by X-rays is not likely to exceed 30\%, heating
of the IGM up to temperatures of $\sim$~few~$\times 10^3$~K can be rapidly
obtained \citep{Shull1985}.  This has interesting implications on the
observability of the reionization process at high redshift. In fact, a
powerful probe of the neutral IGM is 21~cm line emission associated with the
spin-flip transition of the ground state of HI. In order to observe such a
line in emission, the gas temperature needs to be higher that the CMB
temperature. X-rays from microquasars provide the most viable and uniformly
distributed heating source available \citep[e.g.][]{ Chen2004, Kuhlen2006}.
The temperature reached by the IGM subject to a soft X-ray background due to
accretion onto BH can be estimated as \citep{Madau2004}
\begin{equation}
\label{eq:xraytemp}
T \approx 75~{\rm K} \left( \frac{f_{\rm heat}}{0.13} \right) \left(
\frac{\rho_{\rm acc}}{10 {\rm \Msun/Mpc^{3}}}\right)
\end{equation}
with comoving accreted mass density $\rho_{\rm acc}$ ($\rho_{\rm acc} = 10
\Msun/\rm{Mpc}^{3}$ at $z\sim 18$ for our results) and the fraction 
$f_{\rm heat}$ of the X-ray energy going into heating the gas,
depending on the ionization fraction (values for $f_{\rm heat}$ range 
from $f_{\rm heat} \approx 0.13$ to $f_{\rm heat}\approx 0.6$, 
\citet{Shull1985}). As we have discusses, $\rho_{\rm acc}$
is typically smaller than  $\rho_{\rm BH}$ as most the black hole 
mass is built up by BH-mergers.

Summing the total accretion rate over the 3-$\sigma$ halos with $M_{halo}
=10^{8-11} \Msun$ as for the ionizing photon production estimate, we see
(Fig.~\ref{fig:xheat}) that heating by BH accretion becomes important in our
model at $z\lesssim 15-18$ (depending on $f_{\rm heat}$), somewhat later than
the redshifts $z\approx 22$ found by \citet{Madau2004}, whose $\rho_{\rm acc}$
already approached a few hundreds $\Msun/{\rm Mpc}^{3}$ at that redshift.
 The temperatures we find for  $z\lesssim 15-18$ is in the range
$300-1000$, which are somewhat lower than those found by \citet{Kuhlen2005}.
X-rays from SN remnants, shocks associated with structure formation and 
Ly-$\alpha$ photons may also aid the IGM heating process.

\section{Conclusions}
\label{sec:concl}

In this paper we have assessed the early growth and evolution of BH seeds
assumed to form with $M_{seed} \sim 10^4 \Msun$ (an optimistic value) as end
product of the very first generation of PopIII stars or by direct collapse
of gas at $z\sim 20-30$. We have used direct SPH simulations with a simple
prescription for BH gas accretion and associated feedback to explore the
early accretion history of seed black holes in halos of mass $10^8 - 10^{11}
\Msun$, representative of the typical 3-4 $\sigma$ halos collapsing at
$20<z<6$. We have focused on constraining the gas accretion history onto the
seed black holes to compare with the widely adopted assumption that at these
times  seed BHs grow at the Eddington rate. The latter is required if such
seed  black holes are to explain the masses associated with $z\sim 6$
quasars  (without postulating a phase of supercritical accretion as invoked
by \citet[e.g.][]{Volonteri2005b}). In addition, Eddington accretion is
generally assumed when estimating the impact of mini-quasar to reionization
and their role heating the IGM. We found that:

\begin{itemize}

\item seed BHs inhabiting halos of $<10^{11} \Msun$ do not reach an
Eddington rate accretion phase. The growth reaches the Eddington rate at
late times only  for halos of $M \ge 10^{11} \Msun$. Due the limited time
spend in an Eddington growth phase it is difficult to explain the occurrence
of the $z\sim 6$ SDSS quasars.

\item feedback from the BHs is very effective impeding BH growth. Unless 
feedback processes operate at these high redshifts differently such that
black hole feedback return $\ll 0.005$ of the rest-mass energy of the
infalling material to the ISM, its effects leads to a significant delay for
the onset of the Eddington growth phase.

\item the main growth of BHs in these objects occurs mainly through BH
mergers.  The final masses of the central most massive black hole scales as
$M_{BH} \sim 5\times 10^6 (M_{halo}/10^{11} \Msun)^{0.78}$. Because critical 
accretion rates are hard to reach in these halos, the final BH masses depend 
on the initial seed mass.

\item miniquasars are not likely to contribute significantly to reionization.

\item X-ray heating resulting from underfed miniquasars is
however likely sufficient to cause significant IGM  heating
by $z\sim 15$.

\end{itemize}

\acknowledgements
The computations reported here were performed at the Astrophysics Theory
Cluster at Carnegie Mellon University and at the Pittsburgh Supercomputer
Center (PSC), a leading edge site in the NSF Shared Cyberinfrastructure.
TDM acknowledges partial support for this project from NSF grant AST-0607819.

\bibliographystyle{astron}
\bibliography{ms}
   
\end{document}